\RequirePackage{fix-cm} % Fix LaTeX2e bugs.
\documentclass[a4paper, twoside, reqno, dvips, 12pt]{amsart}
\usepackage{fixltx2e}   % Fix LaTeX2e bugs.

\usepackage{etex}

\usepackage[latin1]{inputenc}
\usepackage[T1]{fontenc}

\usepackage{esint}
\usepackage{dsfont}
\usepackage{xspace}
\usepackage{amsgen}
\usepackage{amsthm}
\usepackage{amssymb}
\usepackage{amsmath}
\usepackage{wasysym}
\usepackage{upgreek}
\usepackage{amsfonts}
\usepackage{stmaryrd}
\usepackage{mathtools}

\usepackage{relsize}
\usepackage{textcomp}
\usepackage{textgreek}
\usepackage[mathcal, mathscr]{euscript}

\usepackage{mathrsfs}
\DeclareMathAlphabet{\mathscrbf}{OMS}{mdugm}{b}{n}

\usepackage{a4wide}

\usepackage[dvipsnames, table]{xcolor}
\definecolor{bckg}{RGB}{20.8, 20.8, 20.8}
\definecolor{oneblue}{rgb}{0.0, 0.0, 0.85}
\definecolor{Lightblue}{RGB}{214, 214, 214}
\definecolor{bluepigment}{rgb}{0.2, 0.2, 0.6}
\definecolor{charcoal}{rgb}{0.21, 0.27, 0.31}
\definecolor{denimblue}{rgb}{0.08, 0.38, 0.74}
\definecolor{Lightgray}{rgb}{0.89, 0.89, 0.89}
\definecolor{darkgrey}{rgb}{0.273, 0.281, 0.30}
\definecolor{darkelectricblue}{rgb}{0.33, 0.41, 0.47}

\usepackage{multirow}

\usepackage[sort&compress, comma, square, numbers]{natbib}

\usepackage{psfrag}
\usepackage{graphicx}
\usepackage{adjustbox}
\usepackage[tight]{subfigure}
\usepackage{morefloats}
\usepackage{indentfirst}

\usepackage[usenames, dvipsnames, pdf]{pstricks}
\usepackage{epsfig}
\usepackage{pst-grad} % For gradients
\usepackage{pst-plot} % For axes

\usepackage{rotating}
\usepackage{pdflscape}

\usepackage{acronym}
\usepackage{microtype}
\usepackage[labelsep=period,%
            labelfont={bf,sf,color=bluepigment},%
            justification=raggedright]{caption}

\usepackage[perpage, symbol]{footmisc}

\usepackage[colorlinks,
           urlcolor=oneblue,
           linkcolor=denimblue,
           citecolor=NavyBlue,
           bookmarksopen=false,
           pdfpagemode=UseNone,
           pagebackref]{hyperref}

\usepackage[explicit]{titlesec}

\titleformat{\section}[block]
  {\color{NavyBlue}\Large\sffamily\bfseries}
  {}
  {0.0em}
  {\colorbox{bckg!5}{\strut\parbox{\dimexpr\linewidth-2\fboxsep\relax}{\thesection. #1}}}
  [\vspace*{0.33em}]

\titleformat{name=\section,numberless}[block]
  {\color{NavyBlue}\Large\sffamily\bfseries}
  {}
  {0.0em}
  {\colorbox{bckg!5}{\strut\parbox{\dimexpr\linewidth-2\fboxsep\relax}{#1}}}
  [\vspace*{0.33em}]

\titleformat{\subsection}
  {\color{NavyBlue}\large\sffamily\bfseries}
  {}
  {0.0em}
  {\colorbox{bckg!5}{\parbox{\dimexpr\linewidth-2\fboxsep\relax}{\thesubsection. #1}}}
  [\vspace*{0.33em}]

\titleformat{name=\subsection,numberless}
  {\color{NavyBlue}\Large\sffamily\bfseries}
  {}
  {0em}
  {\colorbox{bckg!5}{\parbox{\dimexpr\linewidth-2\fboxsep\relax}{#1}}}
  [\vspace*{0.33em}]

\titleformat{\subsubsection}
  {\color{bluepigment}\sffamily\normalsize\bfseries}
  {\thesubsubsection}
  {0.5em}
  {#1}
  [\vspace*{0.33em}]

\titleformat{\paragraph}[runin]
  {\color{bluepigment}\sffamily\small\bfseries}
  {}
  {0em}
  {#1}

\titlespacing{\section}{0.0em}{1.5em plus 2pt minus 2pt}%
{1.0em plus 2pt minus 2pt}[0em]
\titlespacing{\subsection}{0.5em}{1.5em plus 2pt minus 2pt}%
{1.0em}[0em]
\titlespacing{\subsubsection}{0.5em}{1.5em plus 2pt minus 2pt}%
{1.0em plus 2pt minus 2pt}[0em]

\usepackage{titletoc}

\contentsmargin{0.5em}
\newlength{\tocsep} 
\titlecontents{section}[\tocsep]
  {\addvspace{10pt}\bfseries\sffamily}
  {\contentslabel[\thecontentslabel]{\tocsep}}
  {}
  {\ \titlerule*[0.75pc]{.}\ \thecontentspage}
  []
\titlecontents{subsection}[\tocsep]
  {\addvspace{8pt}\sffamily}
  {\contentslabel[\thecontentslabel]{\tocsep}}
  {}
  {\ \titlerule*[0.5pc]{.}\ \thecontentspage}
  []
\titlecontents*{subsubsection}[\tocsep]
  {\addvspace{2pt}\footnotesize\sffamily}
  {}
  {}
  {\ \titlerule*[0.35pc]{.}\ \thecontentspage}
  [\\*]

\makeatletter
\def\@setauthors{%
  \begingroup
  \def\thanks{\protect\thanks@warning}%
  \trivlist
  \centering\footnotesize \@topsep30\p@\relax
  \advance\@topsep by -\baselineskip
  \item\relax
  \author@andify\authors
  \def\\{\protect\linebreak}%
  \textsc{\normalsize\textcolor{darkelectricblue}{\authors}}%
  \ifx\@empty\contribs
  \else
    ,\penalty-3 \space \@setcontribs
    \@closetoccontribs
  \fi
  \endtrivlist
  \endgroup
}
\def\@settitle{\begin{center}%
  \baselineskip14\p@\relax
    \bfseries
    \textsc{\Large\textcolor{charcoal}{\@title}}
  \end{center}%
}
\makeatother

\usepackage{enumitem}
\setlist[description]{%
  topsep=30pt,               % space before start / after end of list
  itemsep=5pt,               % space between items
  font={\bfseries\sffamily\color{NavyBlue}}, % if colour is needed
}

\usepackage{fancyhdr}
\usepackage{lastpage}

\newcommand*\Title{\textcolor{bluepigment}{Turbidity current velocity}}
\newcommand*\Authors{\textcolor{bluepigment}{V.~Liapidevskii \& D.~Dutykh}}
\newcommand*{\plogo}{\textcolor{gray}{{\texttt{arXiv.org} / \textsc{hal}}}} % Generic publisher logo

\numberwithin{equation}{section}

\newtheorem{remark}{Remark}

\newcommand{\s}{{\sf s}}
\newcommand{\cm}{{\sf cm}}
\newcommand{\g}{\mathbf{g}}

\newcommand{\R}{\mathds{R}}

\newcommand{\ud}{\mathrm{d}}

\newcommand{\ut}{\tilde{\u}}
\newcommand{\Id}{\mathbb{I}}
\newcommand{\D}{\mathcal{D}}
\newcommand{\E}{\mathcal{E}}
\newcommand{\M}{\mathcal{M}}
\newcommand{\Q}{\mathcal{Q}}
\renewcommand{\phi}{\varphi}
\newcommand{\Cc}{\mathcal{C}}
\renewcommand{\beta}{\upbeta}

\renewcommand{\leq}{\leqslant}
\renewcommand{\geq}{\geqslant}
\newcommand{\eps}{\varepsilon}
\newcommand{\rhob}{\bar{\rho}}
\renewcommand{\H}{\mathcal{H}}
\renewcommand{\P}{\mathcal{P}}
\renewcommand{\alpha}{\upalpha}
\newcommand{\x}{\boldsymbol{x}}
\renewcommand{\theta}{\vartheta}
\newcommand{\vO}{\boldsymbol{0}}
\renewcommand{\b}{\boldsymbol{b}}
\renewcommand{\u}{\boldsymbol{u}}
\newcommand{\const}{\mathrm{const}}
\newcommand{\vtheta}{\text{\textomega}}

\newcommand{\Aa}{\mathscr{A}}
\newcommand{\Bb}{\mathscr{B}}

\newcommand{\Fr}{\mathrm{Fr}}
\newcommand{\RE}{\mathrm{Re}}

\newcommand{\cf}{\emph{c.f.\xspace}}
\newcommand{\ie}{\emph{i.e.\xspace}}
\newcommand{\eg}{\emph{e.g.\xspace}}

\renewcommand{\sim}{\thicksim}
\renewcommand{\div}{\grad\scal}

\newcommand{\scal}{\boldsymbol{\cdot}}
\newcommand{\grad}{\boldsymbol{\nabla}}

\newcommand{\abs}[1]{\lvert\, #1\, \rvert}
\newcommand{\sign}{\mathop{\mathrm{sign}}}

\newcommand{\pd}[2]{\frac{\partial #1}{\partial\/ #2}}
\newcommand{\od}[2]{\frac{\mathrm{d} #1}{\mathrm{d}\/#2}}

\newcommand{\odd}[2]{\dfrac{\mathrm{d} #1}{\mathrm{d}\/#2}}
\newcommand{\eqdef}{\mathop{\stackrel{\,\mathrm{def}}{:=}\,}}
\newcommand{\defeq}{\mathop{\stackrel{\,\mathrm{def}}{=:}\,}}

\newcommand{\half}{{\textstyle{1\over2}}}

\usepackage{acronym}
\acrodef{bvp}[BVP]{Boundary Value Problem}
\acrodef{NSWE}{Nonlinear Shallow Water Equations}

\begin{document}

\title[\Title]{On the velocity of turbidity currents over moderate slopes}

\author[V.~Liapidevskii]{Valery~Yu.~Liapidevskii}
\address{\textbf{V.~Liapidevskii:} Novosibirsk State University and Lavrentyev Institute of Hydrodynamics, Siberian Branch of RAS, 15 Av.~Lavrentyev, 630090 Novosibirsk, Russia}
\email{vliapid@mail.ru}
\urladdr{https://www.researchgate.net/profile/V\_Liapidevskii/}

\author[D.~Dutykh]{Denys Dutykh$^*$}
\address{\textbf{D.~Dutykh:} Univ. Grenoble Alpes, Univ. Savoie Mont Blanc, CNRS, LAMA, 73000 Chamb\'ery, France and LAMA, UMR 5127 CNRS, Universit\'e Savoie Mont Blanc, Campus Scientifique, F-73376 Le Bourget-du-Lac Cedex, France}
\email{Denys.Dutykh@univ-smb.fr}
\urladdr{http://www.denys-dutykh.com/}
\thanks{$^*$ Corresponding author}


%%% ----------------------------------------------------------------------- %%%

\begin{titlepage}
\thispagestyle{empty} % Remove page numbering on this page
\noindent
{\Large Valery \textsc{Liapidevskii}}\\
{\it\textcolor{gray}{Lavrentyev Institute of Hydrodynamics, Novosibirsk, Russia}}\\
{\it\textcolor{gray}{Novosibirsk State University, Novosibirsk, Russia}}
\\[0.02\textheight]
{\Large Denys \textsc{Dutykh}}\\
{\it\textcolor{gray}{CNRS--LAMA, Universit\'e Savoie Mont Blanc, France}}
\\[0.08\textheight]

\vspace*{1.1cm}

\colorbox{Lightblue}{
  \parbox[t]{1.0\textwidth}{
    \centering\huge\sc
    \vspace*{0.7cm}
    
    \textcolor{bluepigment}{On the velocity of turbidity currents over moderate slopes}
    
    \vspace*{0.7cm}
  }
}

\vfill % Whitespace between the title block and the publisher

\raggedleft     % Right-align all text
{\large \plogo} % Publisher and logo
\end{titlepage}

%%% ----------------------------------------------------------------------- %%%

\newpage
\thispagestyle{empty} % Remove page numbering on this page
\par\vspace*{\fill}   % Whitespace until the bottom
\begin{flushright} % Right-align all text
{\textcolor{denimblue}{\textsc{Last modified:}} \today}
\end{flushright}

%%% ----------------------------------------------------------------------- %%%

\newpage
\maketitle
\thispagestyle{empty}

%%% ----------------------------------------------------------------------- %%%

\begin{abstract}

In the present article we consider the problem of underwater avalanches propagating over moderate slopes. The main goal of our work is to investigate the avalanche front velocity selection mechanism when it propagates downwards. In particular, we show that the front velocity does not depend univocally on the mass of sediments. This phenomenon is investigated and explained in our study. Moreover, we derive from the first principles a depth-averaged model. Then, we assume that sediments are uniformly distributed along the slope. In this case, they can be entrained into the flow head and a self-sustained regime can be established. One of the main findings of our study is that the avalanche front velocity is not unique due to a hysteresis phenomenon. We attempt to explain this phenomenon using dynamical systems considerations.

\bigskip
\noindent \textbf{\keywordsname:} turbidity currents; density flows; self-similar solutions; head velocity; moderate slopes \\


\end{abstract}

%%% ----------------------------------------------------------------------- %%%

\newpage
\tableofcontents
\thispagestyle{empty}

%%% ----------------------------------------------------------------------- %%%

\newpage
\section{Introduction}

Perhaps, the first serious attempts to observe turbidity currents in natural environments were performed in late 1960's at \textsc{Scripps} Canyon offshore of \textsc{La Jolla}, \textsc{California}. They were reported in \cite{Inman1976}. The same processes take place on somehow smaller scales in the lakes as well. The first recognition of the r\^ole of turbidity currents in limnology goes back to \textsc{Forel} (1885) \cite{Forel1885}, who conjectured that a sub-aqueous canyon in Lake \textsc{Geneva} had been created by underflows from the \textsc{Rh\^one} river. This process was investigated experimentally in \cite{Kuenen1938}, which is the first experimental study of density currents to our knowledge. Thus, \textsc{Kuenen} (1938) \cite{Kuenen1938} recognized scientifically their potential importance in the transport of sediments. Turbidity currents and underwater landslides are the principal natural mechanisms of sediment transport from shallow to deep waters. Transport distances range from hundred meters to hundred kilometers for Ocean bottoms. This distance is generally referred as the \emph{run-out}. Erosion and deposition by turbidity currents are responsible for numerous features observable on the Ocean bottom. Moreover, we know today that many hydrocarbon reservoirs consist of turbidity current deposits.

The present study was conducted at the University \textsc{Savoie Mont Blanc}'s Campus near Lake \textsc{Le Bourget}, the biggest natural lake in \textsc{France}. For instance, the sedimentary depositions in Alpine lakes have been studied recently \cite{Beck1996, Chapron1999, Chapron2004}. In the same time the water circulation and currents in the Lake \textsc{L\'eman} (also known as Lake of \textsc{Geneva}) have been measured recently using a submarine \cite{Fer2002, Fer2002a, Fer2002b}. The results presented in this study could be applied to some aspects of the limnology as well where the stratified density flows may appear. In natural environments, sub-aqueous sediment gravity flow domain can be conventionally divided into three dynamically distinct regions in the evolution of a downslope current along a slope:
\begin{enumerate}
  \item Source region, where the flow is originated
  \item Transfer region, where the flow accelerates
  \item Deposition region, where the flow decelerates and suspended sediments settle down on the bed.
\end{enumerate}
The gravity current can be divided geometrically into the flow \emph{head}, \emph{body} and \emph{tail}. The head is shaped as an ellipse and is generally thicker than the body. In the present study we are mainly interested in the flow head modelling, where the most intensive mixing processes take place. Consequently, it influences the whole flow dynamics. The most advanced point of the flow head is called the front or nose.

Let us briefly describe the typology of turbidity density currents. The so-called \emph{overflows}\footnote{Overflows are not considered here, but they can be readily described by the proposed model with uneven bathymetries.} are steady or unsteady flows of stratified fluid around underwater obstacles. Usually, one specifies the heavy fluid release upstream the obstacle as a boundary condition. The main modelling problems are the parametrization of the entrainment rate and determination of the velocity of the density fronts. \emph{Thermals} result from a finite heavy fluid mass release into the ambient fluid domain. In laboratory conditions thermals are realized as lock-exchange flows over slopes. The evolution of thermals is generally composed of two distinct phases: (\textit{i}) front acceleration and (\textit{ii}) front deceleration phase. The characteristic property of thermals over rather large slopes is that their shape is well described by appropriate self-similar solutions. Moreover, this regime can be even described by simple \emph{box} (or $0$D) models, where the flow shape is pre-supposed and the evolution of the main flow characteristics is described by Ordinary Differential Equations (ODEs) in time (see \eg~the modelling part in \cite{Rastello2004}). Finally, the \emph{avalanches} are self-sustained density flows. Their sustention mechanism is the entrainment of sedimentary deposits (\eg snow in mountains or sediments under the water) into the flow by turbulence effects. This type of flows lacks laboratory experiments and real-world observation due to their large scales and unforeseeable occurrence. At least, we can say that there is not enough data to infer the dependence of the front velocity on the mass of entrained deposits and the present article is entirely devoted to this question. For instance, in Section~\ref{sec:unsteady} we demonstrate that this dependence turns out to be non-unique. In the following Section~\ref{sec:struct} we try to explain this phenomenon based on the dynamical systems considerations describing travelling wave solutions.

The models employed to study these phenomena range from $0$D (box-type models\footnote{We note that box models have also been used in the context of non-canonical releases, \cf \cite{Zgheib2015}.}) to $3$D (for direct numerical simulation). We stay at the intermediate level of $1$DH modelling, where the second (\ie~vertical) dimension is ruled out using the hydrostatic approximation. Thus, we calculate the evolution along the slope (the horizontal dimension) and in time. In order to improve the model accuracy, several layers might be introduced. For instance, we consider one and two-layer models. In deep channels, one may neglect the motion of the ambient fluid. This allows to simplify the model to a simple single layer nonlinear shallow water equations with mixing\footnote{The mixing terms refer here to the mass and momentum transfer processes to take into account the presence of other layers even if they are not simulated.}. Historically, such depth-averaged models have been extensively used (see \eg \cite{Parker1986}). The main difference between the model proposed in \cite{Parker1986} and the one we use (\cf \cite{Liapidevskii2018}) lies in the fact that our model is more self-contained in the sense that it necessitates no closure relations and fewer modelling parameters to be specified.

The present manuscript is a direct continuation of our previous study \cite{Liapidevskii2018} where the complete derivation and several validations of the mathematical model we employ here can be found. In the present article we investigate further the so-called travelling and self-similar solutions. The question that we address here is the velocity selection mechanism of the underwater avalanche front. In particular, we show using direct numerical simulations that the head front velocity is \emph{practically non-unique} (\ie~small changes of parameters may induce finite changes in the front speed) due to the hysteresis phenomenon. Mathematically speaking, this situation corresponds to the absence of the continuous dependence on certain parameters. We try to explain here the velocity selection mechanism using the standard phase space analysis of the dynamical system, which describes travelling wave solutions to our system. As the main tool in our study, we use a two-layer shallow water system proposed in \cite{Liapidevskii2004} and thoroughly derived, and validated in \cite{Liapidevskii2018}, which includes the entrainment rate of the ambient fluid into the underwater turbidity current. Other particularities of this model can be summarized as follows:
\begin{itemize}
  \item Turbidity current possesses enough energy to absorb and assimilate all the sediments located in the bottom layer,
  \item Governing equations are derived for finite values of the sedimentation velocity,
  \item The entrainment rate of the ambient fluid is determined by the model automatically along with other important parameters,
  \item The model is able to represent at least two types of steady or unsteady density currents:
  \begin{enumerate}
    \item Thermals,
    \item Self-sustained avalanches.
  \end{enumerate}
\end{itemize}
More precisely, we perform a detailed numerical study of unsteady density currents of these two types. The structure of travelling waves is studied in deep channels (so that the ambient fluid can be assumed to be unbounded). Finally, we formulate the avalanche front velocity selection rule for several flow regimes. We discuss also the so-called \emph{hysteresis} phenomenon, which complicates our task. Namely, the front velocity turns out to be dependent on the initiation conditions and, more generally, on the `path' leading the flow to a particular propagation regime.

The present study is organized as follows. The mathematical model is briefly described in Section~\ref{sec:model} (the complete derivation can be found in Appendix~\ref{app:der}). Some special classes of solutions are studied in Section~\ref{sec:spec}. The main conclusions and perspectives of our study are outlined in Section~\ref{sec:disc}.

%%% ----------------------------------------------------------------------- %%%

\section{Mathematical model}
\label{sec:model}

\begin{figure}
  \centering
  \includegraphics[width=1.0\textwidth]{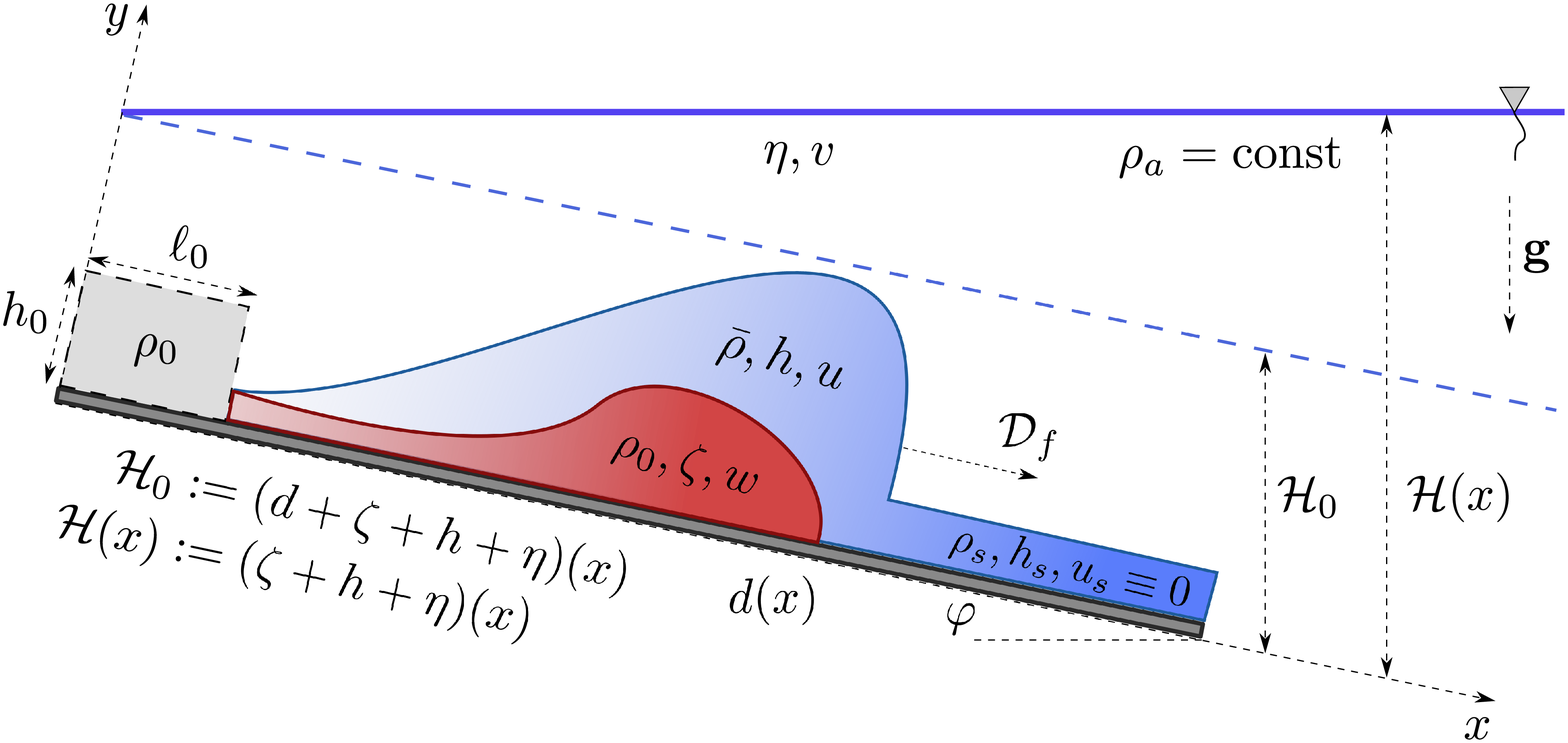}
  \caption{\small\em Sketch of the physical domain and the flow configuration. Various notations and variables are introduced and explained in the text. The bottom layer shown in red on this picture might be included into consideration to describe some special classes of flows such as exchange flows.}
  \label{fig:sketch}
\end{figure}

In this study we consider the physical situation schematically depicted in Figure~\ref{fig:sketch}. The three-layer system is derived in Appendix~\ref{app:der}, where the bottom layer (shown in red) is taken into account. Below we present a reduced system, which is actually used in our investigations. The bottom layer should be taken into account to describe certain classes of flows such as lock-exchange ones.

The depth-averaged governing equations, which describe the mixing layer dynamics are given by the following system of partial differential equations:
\begin{align}\label{eq:main1}
  h_{\,t}\ +\ \bigl[\,h\,u\,\bigr]_{\,x}\ &=\ \sigma\,q\,, \\
  (b\,h)_{\,t}\ +\ \bigl[\,b\,h\,u\,\bigr]_{\,x}\ &=\ 0\,, \label{eq:main2}\\
  v_{\,t}\ +\ \bigl[\,\half\,v^{\,2}\ +\ p\,\bigr]_{\,x}\ &=\ 0\,, \label{eq:main3}
\end{align}
\begin{equation}\label{eq:main4}
  (h\,u\ +\ \eta\,v)_{\,t}\ +\ \Bigl[\,h\,u^{\,2}\ +\ \half\,b\,h^{\,2}\ +\ \H\,p\,\Bigr]_{\,x}\ =\ -\,(p\ +\ b\,h)\,d_{\,x}\ -\ u_{\,\ast}^{\,2}\,,
\end{equation}
\begin{multline}\label{eq:main5}
  \bigl(h\,(u^{\,2} + q^{\,2}) + \eta\,v^{\,2} + b\,h^{\,2}\bigr)_{\,t}\ +\ \bigl[\,h\,u\,(u^{\,2} + q^{\,2}) + \eta\,v^{\,3} + 2\,p\,\Q + 2\,b\,h^{\,2}\,u\,\bigr]_{\,x}\\
   =\ -\,2\,b\,h\,u\,d_{\,x}\ -\ \kappa\,\abs{q}^{\,3}\ -\ 2\,u\,u_{\,\ast}^{\,2}\,,
\end{multline}
The physical sense of various variables is explained in Table~\ref{tab:defs}. The detailed derivation of Equations~\eqref{eq:main1} -- \eqref{eq:main5} can be found in Appendix~\ref{app:der}. This model was extensively tested and validated in our previous study \cite{Liapidevskii2018}.

System \eqref{eq:main1} -- \eqref{eq:main5} is the main model appropriate for the modelling unsteady density currents of the underwater thermal and avalanche types. Similarly to System~\eqref{eq:sys1} -- \eqref{eq:sys7} we can compute two following differential consequences (\cf~Equations~\eqref{eq:6a}, \eqref{eq:6b}):
\begin{equation*}
  b_{\,t}\ +\ u\,b_{\,x}\ =\ -\,\frac{\sigma\,q\,b}{h}\,, \\
\end{equation*}
\begin{equation*}
  q_{\,t}\ +\ u\,q_{\,x}\ =\ \frac{\sigma}{2\,h}\,\bigl((u\ -\ v)^{\,2}\ -\ (1\ +\ \delta)\,q^{\,2}\ -\ b\,h\bigr)\,,
\end{equation*}
where the parameter $\delta\ \eqdef\ \dfrac{\kappa}{\sigma}$ is introduced for the sake of convenience. These two non-conservative equations show that for the two-layer system the multiple contact characteristics $\od{x}{t}\ =\ \nu_{\,0}^{\,(1,\,2)}\ \eqdef\ u$ are preserved. Two remaining characteristics coincide with characteristics of classical two-layer shallow water equations \cite{Liapidevskii2000} and they can be computed analytically:
\begin{equation*}
  \nu_{\,\pm}^{\,(3,\,4)}\ \eqdef\ \frac{(\H\ -\ h)\,u\ +\ h\,v}{\H}\ \pm\ \frac{\sqrt{h\,(\H\ -\ h)\,(b\,\H\ -\ (u\ -\ v)^{\,2})}}{\H}\,.
\end{equation*}
From the last formula it follows that System \eqref{eq:main1} -- \eqref{eq:main5} is hyperbolic if the velocity difference in two layers satisfy the inequality:
\begin{equation}\label{eq:10}
  (u\ -\ v)^{\,2}\ \leq\ b\,\H\,.
\end{equation}
In inclined channels of finite depth, the velocity of the bottom layer may achieve the critical value beyond which the hyperbolicity condition \eqref{eq:10} is not verified. In this case Equations \eqref{eq:main1} -- \eqref{eq:main5} are not applicable anymore since they will provide unstable solutions in the elliptic regime (also known as \textsc{Hadamard}'s instability). In this work we use the term `moderate slopes' to designate a class of channels for which the thickness of the upper (quiescent fluid) layer is sufficient to ensure the model \eqref{eq:main1} -- \eqref{eq:main5} strict hyperbolicity.

\begin{remark}
The numerical resolution of Systems \eqref{eq:cons1} -- \eqref{eq:cons4} and \eqref{eq:main1} -- \eqref{eq:main5} in the domain of hyperbolicity is quite straightforward, since equations are written in the suitable conservative form of quasi-linear balance laws (see \eg~\cite{LeVeque1992, Kroner1997}). In the present study we employ the celebrated \textsc{Godunov} scheme in our simulations \cite{Godunov1959, Godunov1999}. Its main advantages include very low numerical dissipation due to the exact solution of local \textsc{Riemann} problems at cell interfaces. The explicit \textsc{Euler} scheme is used in time discretization. It preserves the monotonicity of solutions and has several other good properties that we expect from a robust numerical method \cite{Amadori2004, LeVeque1985}.
\end{remark}

%%% ----------------------------------------------------------------------- %%%

\section{Model validation}
\label{sec:valid}

In this Section we check the applicability of a simple one-layer model \eqref{eq:main1} -- \eqref{eq:main5}. We apply it to simulate the sediments entrainment process by density currents over rather moderate slopes. Numerical solutions will be compared with experimental data reported in \cite{Rastello2004} as well as with exact special solutions of travelling wave type. Moreover, we shall show below that under certain initial conditions the numerical solution will tend asymptotically to self-similar solutions described below.

Sketch of the numerical and laboratory experiment is shown in Figure~\ref{fig:sketch}. The initial and boundary conditions are rather standard as well. Initially, at $t\ =\ 0\,$, the distribution of evolutionary quantities is given on the computational domain $[\,0,\,\ell\,]\,$:
\begin{multline*}
  h\,(x,\,0)\ =\ h_{\,0}\,(x)\,, \qquad
  u\,(x,\,0)\ =\ u_{\,0}\,(x)\,, \qquad
  v\,(x,\,0)\ =\ v_{\,0}\,(x)\,, \\
  m\,(x,\,0)\ =\ m_{\,0}\,(x)\,, \qquad
  q\,(x,\,0)\ =\ q_{\,0}\,(x)\,.
\end{multline*}
On the right boundary we set wall boundary conditions for simplicity\footnote{In any case, the simulation stops before the mass reaches the right boundary. So, the influence of the right boundary condition on presented numerical results is completely negligible.}. On the left extremity of the computational domain the boundary conditions depend on the flow regime in the vicinity of the left boundary. If the flow is supercritical, we impose \textsc{Cauchy}'s data. Otherwise, we impose a wall boundary condition as well. The case we study will be described in more details below.

According to the experiments of \textsc{Rastello} \& \textsc{Hopfinger} (2004) \cite{Rastello2004} we use the following configuration of the numerical tank. The channel length $\ell$ is equal to $200\;\cm\,$. On boundaries we impose wall boundary conditions. A dense fluid of buoyancy $b_{\,\ell}$ fills an initially closed container of the length of $20\;\cm$ and of the height $h_{\,\ell}\,$. This configuration corresponds to the classical lock-exchange experiment. At the distance of $50\;\cm$ from the removable door, the slope is covered by initially motionless sediment layer of the height $h_{\,s}\,$. The ``mass'' of sediments is $m_{\,s}\ =\ b_{\,s}\cdot h_{\,s}\,$. The length of sediments layer is $100\;\cm\,$. During the propagation of the heavy fluid head all these sediments were entrained into the flow. In some experiments the sediment layer was absent, \ie $m_{\,s}\ \equiv\ 0\,$. In this case the flow simply propagates over the rigid inclined bottom as a thermal. In order to avoid the degeneration of certain equations, in numerical experiments we slightly cover the whole slope with a micro-layer of sediments with mass $m_{\,s}^{\,\circ}\ >\ 0\,$, such that
\begin{equation*}
  m_{\,s}^{\,\circ}\ \ll\ m_{\,s} \qquad \mbox{and} \qquad
  m_{\,s}^{\,\circ}\ \ll\ m_{\,\ell}\,.
\end{equation*}
This micro-layer starts at the removable wall and extends to the right boundary of the computational domain. In this way we can compute with the same numerical code the transient propagation of thermals ($m_{\,s}\ \equiv\ 0$) and density currents lifting up the sediments ($m_{\,s}\ >\ 0$). The values of all physical parameters used in our computations are reported in Table~\ref{tab:params0}. The total channel depth $\H$ is chosen to be sufficiently large so that this parameter do not influence anymore the computational results. In all numerical simulations reported below we used the following values of these parameters:
\begin{equation*}
  \sigma\ =\ 0.15\,, \qquad \delta\ =\ 0\,, \qquad c_{\,w}\ =\ 0\,.
\end{equation*}

\begin{table}
  \centering
  \begin{tabular}{c|c|c}
  \hline\hline
  \textit{Parameter} & \textit{Experiment 1} & \textit{Experiment 2} \\
  \hline\hline
  Slope angle, $\varphi$ & $32^{\,\circ}$ & $45^{\,\circ}$ \\
  Container height, $h_{\,\ell}$ & $6.5\ \cm$ & $20\ \cm$ \\
  Heavy fluid buoyancy, $b_{\,\ell}$ & $19\ \dfrac{\cm}{s^2}$ & $1\ \dfrac{\cm}{s^2}$ \vspace*{0.2em} \\
  Sediment deposit height, $h_{\,s}$ & $0.2\ \cm$ & $2.0\ \cm$ \\
  Minimal sediment ``mass'', $m_{\,s}^{\,\circ}$ & $10^{-3}\;\dfrac{\cm^2}{\s^2}$ & $10^{-3}\;\dfrac{\cm^2}{\s^2}$ \vspace*{0.2em} \\
  Sediment ``mass'', $m_{\,s}$ & $20\;\dfrac{\cm^2}{\s^2}$ & $3\;\dfrac{\cm^2}{\s^2}$ \vspace*{0.2em} \\
  Final simulation time, $T$ & $10\ \s$ & $20\ \s$ \\
  \hline\hline
  \end{tabular}
  \bigskip
  \caption{\small\em Parameters used in numerical simulations of the experiments from \textsc{Rastello} \& \textsc{Hopfinger} (2004) \cite[Table~2]{Rastello2004}.}
  \label{tab:params0}
\end{table}

\begin{figure}
  \centering
  \includegraphics[width=0.99\textwidth]{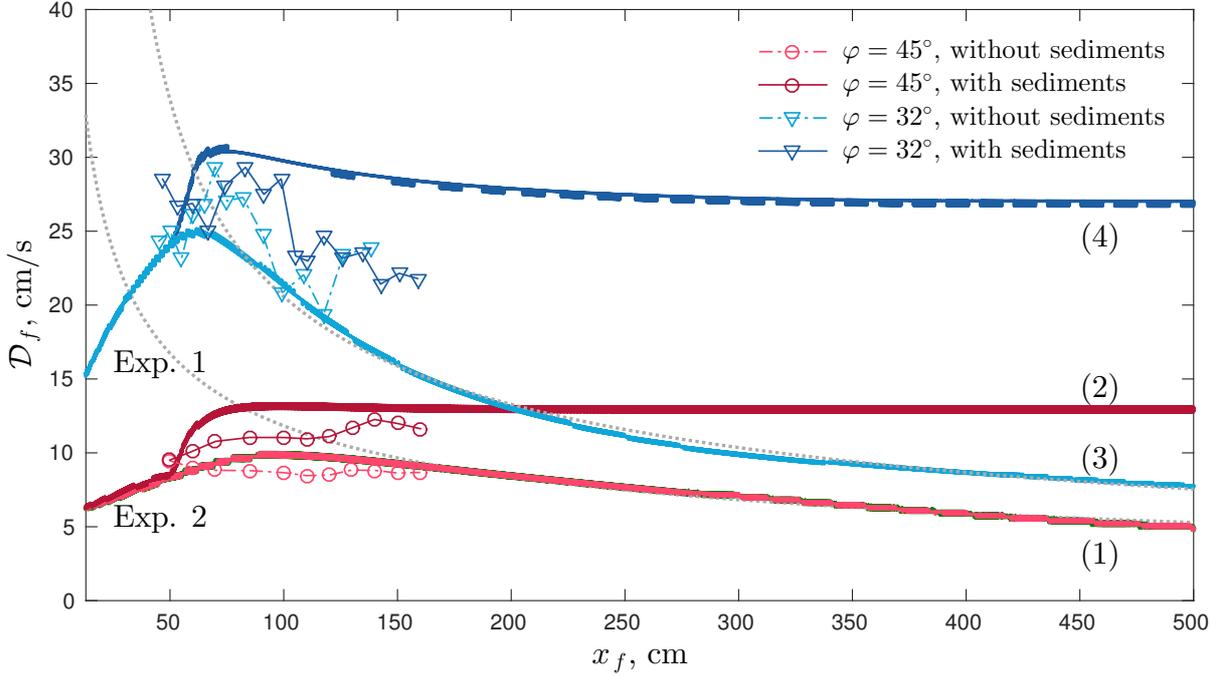}
  \caption{\small\em Evolution of the front velocity $\D_{\,f}$ of laboratory clouds as a function of downstream distance $x_{\,f}\,$. Comparison of the predicted flow head velocity (lines without markers) against the experimental results (lines with markers) from \textsc{Rastello} \& \textsc{Hopfinger} (2004) \cite[Fig.~11]{Rastello2004}. The upper (lower) group of data corresponds to Experiment~1 (2) from Table~\ref{tab:params0}. Dotted lines show the asymptotic regime $\D_{\,f}\ \sim\ \dfrac{1}{\sqrt{x_{\,f}}}$ demonstrated in \cite{Liapidevskii2018}.}
  \label{fig:2n}
\end{figure}

In Figure~\ref{fig:2n} we report the dependence of the flow head celerity $\D_{\,\mathrm{f}}$ on the distance $x_{\,\mathrm{f}}$ traveled by the front. Notice that all quantities are shown in dimensional (physical) variables. The differences between experiments $1$ and $2$ are highlighted in Table~\ref{tab:params0}. Various symbols ($\bigcirc$ and $\bigtriangledown$) correspond to laboratory measurements of the head velocity and are taken from \cite[Fig.~11]{Rastello2004}. Solid and dash-dotted lines (with $\bigcirc$ and $\bigtriangledown$) indicate the experiments with ($m_{\,s}\ >\ 0$) and without ($m_{\,s}\ =\ 0$) sediments correspondingly. The presence of sediments before the head front yields, obviously, to the initial quite visible additional acceleration of the flow. In Figure~\ref{fig:2n} we show also the long time behaviour of the front velocity from the Experiments~$1$ \& $2\,$. In order to perform this simulation we increased the computational domain to $500\;\cm\,$. In numerical experiments with the sediment layer ($m_{\,s}\ >\ 0$) we extend this layer up to the right end of the channel to reveal completely the asymptotic behaviour of density fronts. The excellent agreement with theoretically established in \cite{Liapidevskii2018} asymptotics $\D_{\,\mathrm{f}}\ \sim\ \dfrac{1}{\sqrt{x_{\,\mathrm{f}}}}$ (dotted lines) can be observed in this Figure~\ref{fig:2n} for the case $m_{\,s}\ =\ 0$ (curves $1$ and $3$ show thermals evolution). It validates both the numerics and the underlying theoretical arguments. For the case $m_{\,s}\ >\ 0$ we see that the front celerity $\D_{\,\mathrm{f}}$ reaches the constant asymptotic value $\D_{\,\mathrm{asym}}\ =\ \Cc(\phi)\cdot \sqrt{m_{\,s}}$ and numerical computations (curves $2$ and $4$) show that
\begin{equation}\label{eq:31}
  \Cc\,(32^{\,\circ})\ \approx\ 6.32\,, \qquad
  \Cc\,(45^{\,\circ})\ \approx\ 7.5\,.
\end{equation}
These calculated values will be used below for the validation of the theoretical model.

%%% ----------------------------------------------------------------------- %%%

\section{Special solutions}
\label{sec:spec}

In this Section we study some properties of solutions to System \eqref{eq:main1} -- \eqref{eq:main5} in the case where the upper layer depth $\H$ is much bigger than the mixing layer depth, \ie~$h\ \ll\ \H\,$. If additionally the upper layer is at rest, then $v\ \equiv\ 0\,$, $p\ \equiv\ 0$ and the system of governing equations is further simplified (we omit here the friction terms):
\begin{align}\label{eq:11a}
  h_{\,t}\ +\ \bigl[\,h\,u\,\bigr]_{\,x}\ &=\ \sigma\,q\,, \\
  (b\,h)_{\,t}\ +\ \bigl[\,b\,h\,u\,\bigr]_{\,x}\ &=\ 0\,,\label{eq:11b} \\
  (h\,u)_{\,t}\ +\ \Bigl[\,h\,u^{\,2}\ +\ \half\,b\,h^{\,2}\,\Bigr]_{\,x}\ &=\ -\,b\,h\,d_{\,x}\,,\label{eq:11c} \\
  \bigl(h\,(u^{\,2} + q^{\,2}\ +\ b\,h)\bigr)_{\,t}\ +\ \bigl[\,h\,u\,(u^{\,2} + q^{\,2}) + 2\,b\,h\,\bigr]_{\,x}\ &=\ -\,2\,b\,h\,u\,d_{\,x}\ -\ \kappa\,\abs{q}^{\,3}\,. \label{eq:11d}
\end{align}
This system turns out to be a generalization of one layer shallow water equations \cite{Liapidevskii2000, Liapidevskii2018}. Its characteristics can be readily computed:
\begin{equation*}
  \nu^{\,0}_{\,1,\,2}\ =\ u\,, \qquad
  \nu^{\,\pm}_{\,3,\,4}\ =\ u\ \pm\ \sqrt{b\,h}\,.
\end{equation*}
In other words, System \eqref{eq:11a} -- \eqref{eq:11d} is hyperbolic provided that $h\ >\ 0\,$. Properties of this system regarding three types of density currents were extensively studied in \cite{Liapidevskii2018}. Here we focus on the possibility to determine the wave front velocity of an underwater avalanche by constructing some specific exact solutions. In this way, we are naturally lead to consider two types of special solutions:
\begin{itemize}
  \item Travelling waves,
  \item Similarity solutions.
\end{itemize}
They are going to be studied in the remainder of this Section.

%%% ----------------------------------------------------------------------- %%%

\subsection{Travelling waves}
\label{sec:tw}

Consider again System \eqref{eq:11a} -- \eqref{eq:11d}, which describes general density currents over moderate slopes. In this Section we consider the situation where the slope is constant, \ie~$\alpha\ \eqdef\ -d_{\,x}\ =\ \const$. Moreover, we are looking for solutions of the form\footnote{This form will be also referred to as the travelling wave ansatz.}:
\begin{equation}\label{eq:twan}
  h\,(t,\,x)\ \equiv\ h\,(\xi)\,, \quad
  m\,(t,\,x)\ \equiv\ m\,(\xi)\,, \quad
  u\,(t,\,x)\ \equiv\ u\,(\xi)\,, \quad
  q\,(t,\,x)\ \equiv\ q\,(\xi)\,,
\end{equation}
where $\xi\ \eqdef\ x\ -\ \D\,t\,$, $\D\ >\ 0$ is the front velocity and $m\ =\ b\cdot h\,$. For this type of solutions, the speed $\D$ is to be determined from some additional conditions along with solution profiles $h\,(\xi)\,$, $m\,(\xi)\,$, $u\,(\xi)$ and $q\,(\xi)\,$. After substituting the travelling wave ansatz \eqref{eq:twan} into System \eqref{eq:11a} -- \eqref{eq:11d}, we obtain the following system of ODEs:
\begin{align}\label{eq:tw1}
  \bigl(u\ -\ \D\bigr)\,h^{\,\prime}\ +\ h\,u^{\,\prime}\ &=\ \sigma\,q\,, \\
  \bigl(u\ -\ \D\bigr)\,m^{\,\prime}\ +\ m\,u^{\,\prime}\ &=\ 0\,,\label{eq:tw2} \\
  \bigl(u\ -\ \D\bigr)\,u^{\,\prime}\ +\ \half\,m^{\,\prime}\ +\ \frac{m}{2\,h}\;h^{\,\prime}\ &=\ \frac{\alpha\,m\ -\ \sigma\,q\,u}{h}\,,\label{eq:tw3} \\
  \bigl(u\ -\ \D\bigr)\,q^{\,\prime}\ &=\ \frac{\sigma\,\bigl(u^{\,2}\ -\ (1\ +\ \delta)\,q^{\,2}\ -\ m\bigr)}{2\,h}\,,\label{eq:tw4}
\end{align}
where the prime superscript ${}^{\prime}$ denotes the derivative with respect to $\xi$ variable, \ie~$h^{\,\prime}\ \eqdef\ \od{h}{\xi}\,$. By integrating the second equation above, we obtain:
\begin{equation*}
  m\,(\xi)\ =\ m\,\bigl(u\,(\xi)\bigr)\ =\ \frac{\M}{u\ -\ \D}\,, \qquad
  \M\ =\ \const.
\end{equation*}
The rest of equations can be reduced to the following autonomous ODE in the plane $(u,\,q)\,$:
\begin{equation}\label{eq:15}
  \od{q}{u}\ =\ \frac{\sigma}{2}\;\frac{\bigl(u^{\,2}\ -\ (1\ +\ \delta)\,q^{\,2}\ -\ m\bigr)\,\bigl((u\ -\ \D)^{\,2}\ -\ m\bigr)}{(u\ -\ \D)^{\,2}\,\bigl(\alpha\,m\ -\ \sigma\,q\bigl(u\ +\ \frac{m}{2\,(u\ -\ \D)}\bigr)\bigr)}\,.
\end{equation}
Properties of solutions to the last ODE depend on the mutual location in plane $(u,\,q)$ of three curves:
\begin{align*}
  \Gamma_{\,A}\ &\eqdef\ \bigl\{(u,\,q)\;\vert\; u^{\,2}\ -\ (1\ +\ \delta)\,q^{\,2}\ -\ m\ =\ 0\bigr\}\,, \\
  \Gamma_{\,B}\ &\eqdef\ \bigl\{(u,\,q)\;\vert\; \alpha\,m\ -\ \sigma\,q\bigl(u\ +\ \frac{m}{2\,(u\ -\ \D)}\bigr)\ =\ 0\bigr\}\,, \\
  \Gamma_{\,C}\ &\eqdef\ \bigl\{(u,\,q)\;\vert\; u\ =\ u_{\,\mathrm{c}}\bigr\}\,,
\end{align*}
where $u_{\,\mathrm{c}}$ is the critical speed, which solves the following equation:
\begin{equation*}
  \Delta\,(u_{\,\mathrm{c}})\ \eqdef\ (u_{\,\mathrm{c}}\ -\ \D)^{\,2}\ -\ m\,(u_{\,\mathrm{c}})\ =\ 0\,.
\end{equation*}
The last condition provides us the value of the velocity $u$ under which the flow becomes critical in the reference frame moving with speed $\D\,$. The \textsc{Galilean} invariance of governing equations is of capital importance to give sense to such changes of variables \cite{Duran2013}.

The behaviour of travelling wave solutions will be analysed below regarding the problem of sediments entrainment by an underwater avalanche. Moreover, we will highlight the velocity selection mechanism (or criterium, mathematically speaking) for the self-sustained density current front velocity $\D\,$.

%%% ----------------------------------------------------------------------- %%%

\subsection{Similarity solutions}

In order to describe non-stationary processes taking place behind the density current front, we consider the following class of self-similar solutions:
\begin{equation*}
  h\,(t,\,x)\ =\ t^{\,\vtheta\,+\,1}\,\hat{h}\,(\xi)\,, \quad
  u\,(t,\,x)\ =\ t^{\,\vtheta}\,\hat{u}\,(\xi)\,, \quad
  m\,(t,\,x)\ =\ t^{\,2\,\vtheta}\,\hat{m}\,(\xi)\,, \quad
  q\,(t,\,x)\ =\ t^{\,\vtheta}\,\hat{q}\,(\xi)\,,
\end{equation*}
where $\xi\ \eqdef\ \dfrac{x}{t^{\,\vtheta\,+\,1}}\,$, $\vtheta\ =\ \const$. It can be verified that System \eqref{eq:11a} -- \eqref{eq:11d} admits such solutions under the appropriate choice of the exponent $\vtheta\,$. We are going to consider two cases corresponding to two types of flows under consideration.

%%% ----------------------------------------------------------------------- %%%

\subsubsection{Thermals}

If in front of the wave there are no available sediments, then from the mass conservation law we can deduce:
\begin{equation*}
  \int_{\,0}^{\,+\,\infty} m\,(t,\,x)\;\ud x\ =\ \const.
\end{equation*}
From the last condition it follows that $\vtheta\ =\ -\frac{1}{3}\,$. Then, self-similar solutions take the following form:
\begin{multline*}
  h\,(t,\,x)\ =\ t^{\,2/3}\,\hat{h}\,(\xi)\,, \quad
  u\,(t,\,x)\ =\ t^{\,-1/3}\,\hat{u}\,(\xi)\,, \\
  m\,(t,\,x)\ =\ t^{\,-2/3}\,\hat{m}\,(\xi)\,, \quad
  q\,(t,\,x)\ =\ t^{\,-1/3}\,\hat{q}\,(\xi)\,,
\end{multline*}
where $\xi\ \eqdef\ \dfrac{x}{t^{\,2/3}}\,$. This self-similar solution describes the deceleration phase of thermals on large evolution times. This particular solution can be used for the validation of numerical solutions, when they are plotted in the appropriate self-similar variables.

%%% ----------------------------------------------------------------------- %%%

\subsubsection{Heteroclinic solutions}
\label{sec:ans}

In order to construct unsteady solutions heteroclinic to travelling waves (considered in Section~\ref{sec:tw}), we consider a special class of similarity solutions with $\vtheta\ =\ 0\,$, \ie
\begin{equation}\label{eq:ans}
  h\,(t,\,x)\ =\ t\,\hat{h}\,(\xi)\,, \quad
  u\,(t,\,x)\ =\ \hat{u}\,(\xi)\,, \quad
  m\,(t,\,x)\ =\ \hat{m}\,(\xi)\,, \quad
  q\,(t,\,x)\ =\ \hat{q}\,(\xi)\,,
\end{equation}
with $\xi\ \eqdef\ \dfrac{x}{t}\,$. This class of solutions will be used to analyze the asymptotic behaviour of unsteady numerical solutions in the problem of an underwater avalanche propagation taking into account the entrainment of sediments.

%%% ----------------------------------------------------------------------- %%%

\subsection{Validation of similarity solutions}

\begin{figure}
  \centering
  \subfigure[]{\includegraphics[width=0.5\textwidth]{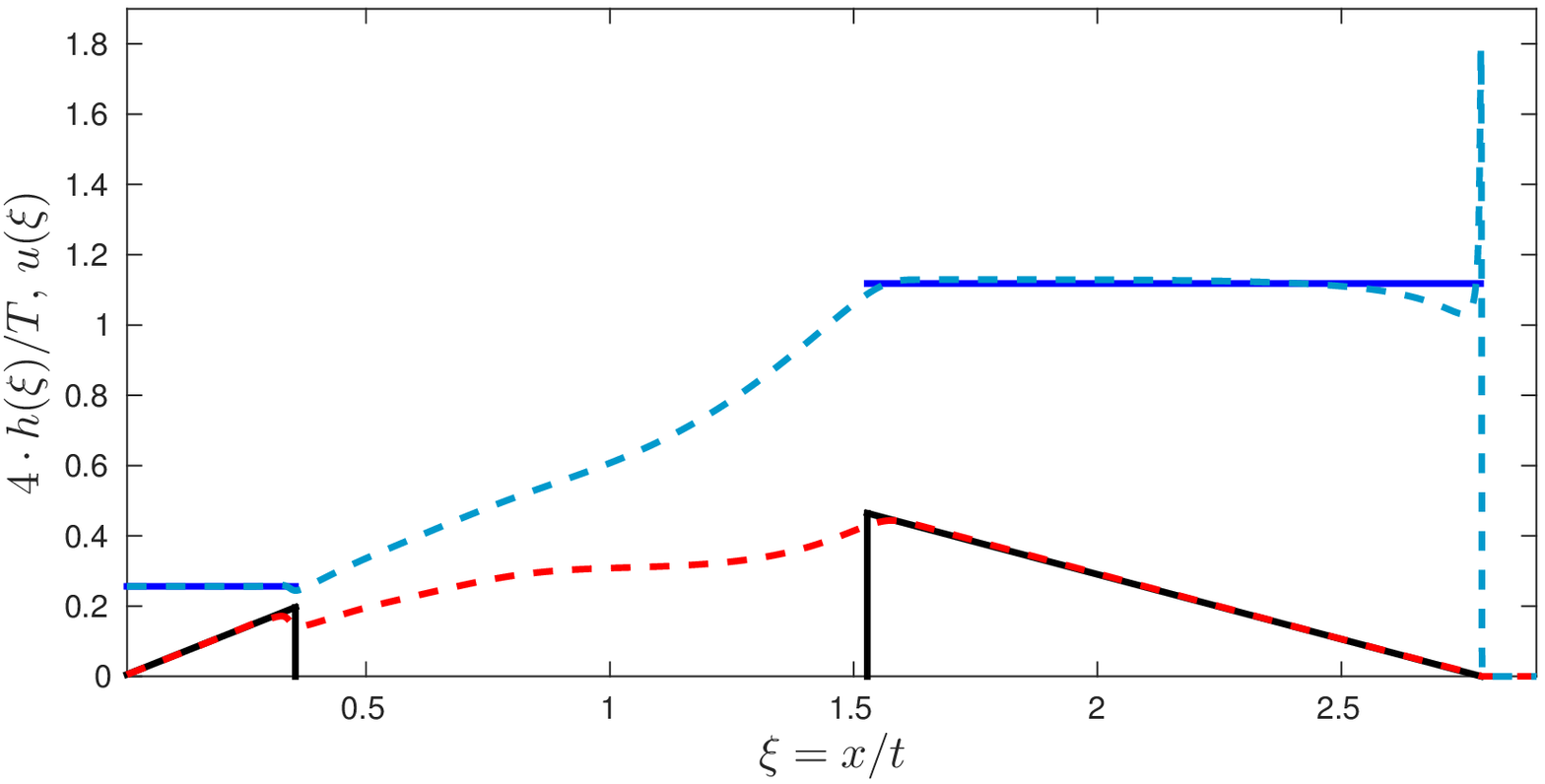}}
  \subfigure[]{\includegraphics[width=0.46\textwidth]{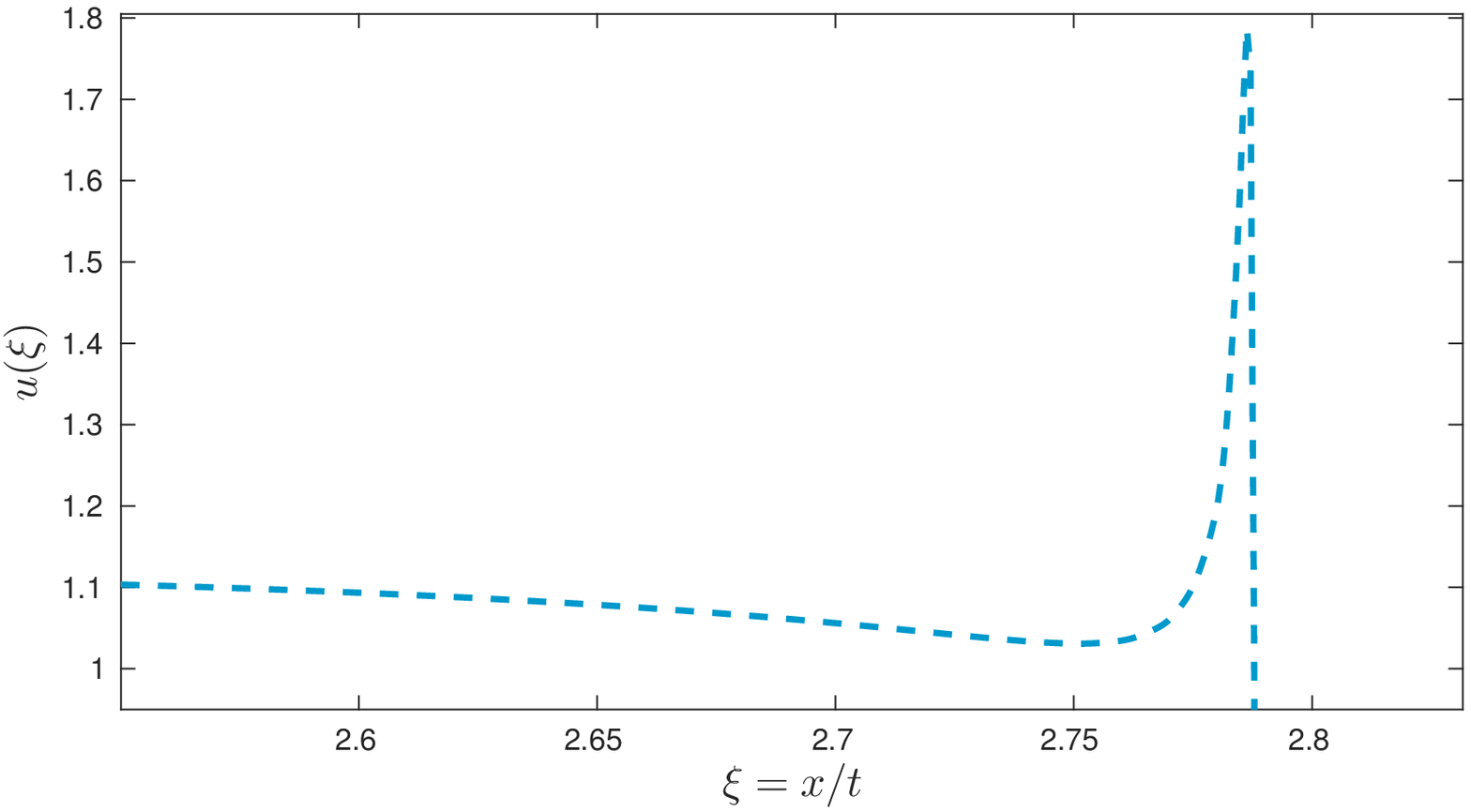}}
  \caption{\small\em Validation of similarity solution by unsteady simulations. Left panel (a): the exact solution is depicted with solid lines (black --- scaled flow height $4\,\frac{h(\xi)}{T}\,$, blue --- $u(\xi)$) and unsteady numerical prediction are shown with dashed lines (red --- layer height, blue --- horizontal speed). Right panel (b): zoom on the jump in the velocity variable at the flow head.}
  \label{fig:3n}
\end{figure}

In Figure~\ref{fig:2n} we demonstrated how unsteady solutions ($m_{\,s}\ >\ 0$ and $m_{\,s}\ =\ 0$) tend to the asymptotic regime. It seems that for a given slope angle $\phi$ this regime is unique. Mathematically this assumption can be translated by the fact that the proportionality coefficient $\Cc\,(\phi)$ is a single-valued function of $\phi\,$. However, we shall see later that it is not the case. In order to demonstrate this property, we consider the same experimental set-up as in Section~\ref{sec:valid}. However, in order to study long time behaviour of the flow, we consider the channel of the length $\ell\ =\ 6\,000\;\cm\,$. The bottom slope is taken to be $\phi\ =\ 45^{\circ}\,$. Starting from $x\ =\ 20\;\cm$ and until the channel end a uniform ($h_{\,s}\ \equiv\ 0.1\;\cm$) motionless ($u_{\,s}\ \equiv\ 0\,$, $q_{\,s}\ \equiv\ 0$) layer of sediments of mass $m_{\,s}\ \equiv\ 0.1\;\frac{\cm^2}{\s^2}$ is located. Sediment particles start to move when the flow head passes and the layer depth increases to $h\ =\ 1.1\,h_{\,s}\,$. In order to initiate a \emph{substantial} density flow in the channel after releasing the heavy fluid, it is sufficient to fill the reservoir located for $x\ \in\ [\,0,\,20\,]$ with the following heavy fluid parameters:
\begin{equation*}
  m_{\,\ell}\ =\ m_{\,s}\,, \qquad
  h_{\,\ell}\ =\ 1.2\,h_{\,s}\,, \qquad
  u_{\,\ell}\ =\ 0\,, \qquad
  q_{\,\ell}\ =\ 0\,.
\end{equation*}
For $t\ \gtrsim\ 1000\;\s$ the numerical solution enters into the asymptotic regime and for all subsequent times the flow picture in self-similar variables does not change. The numerical solution for $T\ =\ 2000\;\s$ is shown in Figure~\ref{fig:3n}. The dashed lines on the left panel (\textit{a}) show the theoretical distributions of the scaled wave height $4\;\frac{h\,(\xi)}{T}$ (red line) along with the velocity profile $u\,(\xi)$ (blue line). In numerical simulations we observed the flow head propagation speed $\D_{\,\mathrm{num}}\ \approx\ 2.79\;\frac{\cm}{\s}\,$. We remark that behind the wave front there is a sufficiently long zone $1.53\ \approx\ \xi_{\,1}\ <\ \xi\ <\ \D_{\,\mathrm{num}}\,$, where the solutions can be very well approximated by linear functions. In this area the wave profile is linear and the velocity is constant. A zoom on the velocity peak at the flow head is shown in the right panel of Figure~\ref{fig:3n}(\textit{b}). Its structure is quite smooth, when the flow is resolved to high resolution.

If at the left boundary of the channel we maintain the constant mass flux $\M_{\,\ell}\ \eqdef\ u_{\,\ell}\cdot m_{\,\ell}$ (we take $\M_{\,\ell}\ \equiv\ 2.5\times 10^{-3}\;\frac{\cm^3}{\s^3}$), then, in the neighbourhood of the left boundary another zone with the linear wave profile and constant velocity is formed. In our numerical simulations this zone occupies the interval $0\ <\ \xi\ <\ \xi_{\,0}\ \approx\ 0.355\,$. The flow in this area is supercritical ($u^{\,2}\ >\ m$). Consequently, on the left boundary $x\ =\ 0$ we have to specify all evolution variables:
\begin{equation*}
  h_{\,\ell}\ =\ 0.005\;\cm\,, \qquad
  u_{\,\ell}\ =\ 0.5\;\frac{\cm}{\s}\,, \qquad
  m_{\,\ell}\ =\ 0.005\;\frac{\cm^2}{\s^2}\,, \qquad
  q_{\,\ell}\ =\ 0\;\frac{\cm}{\s}\,.
\end{equation*}
For $\xi_{\,1}\ <\ \xi\ <\ \D_{\,\mathrm{num}}$ the flow is supercritical in the frame of reference moving with the flow head, \ie~$(\D_{\,\mathrm{num}}\ -\ u)^2\ >\ m\,$. That is why behind the wave front the analogue of \textsc{Chapman}--\textsc{Jouguet} conditions is not verified \cite{Fickett1979}.

Stationary solutions to System \eqref{eq:main1} -- \eqref{eq:main5} of the form
\begin{equation*}
  u\ \equiv\ u_{\,j}\,, \qquad
  m\ \equiv\ m_{\,j}\,, \qquad
  q\ \equiv\ q_{\,j}\,, \qquad
  h\ \equiv\ \lambda_{\,j}\cdot x\,
\end{equation*}
can be easily constructed. These solutions can be used to describe the flow on the interval $(0,\,\xi_{\,0})\,$. Here the value of $\xi_{\,0}$ coincides with the characteristics, \ie~$\xi_{\,0}\ =\ u_{\,j}\ +\ \sqrt{m_{\,j}}\,$. The corresponding solutions on this interval are shown in Figure~\ref{fig:3n} with solid lines. On the other hand, let us consider an exact travelling wave solution of the following form:
\begin{equation*}
  h\,(t,\,x)\ =\ t\,\lambda_{\,\mathrm{f}}\,(\D\ -\ \xi)\,, \qquad
  u\,(t,\,x)\ =\ u_{\,\mathrm{f}}\,, \qquad
  m\,(t,\,x)\ =\ m_{\,\mathrm{f}}\,, \qquad
  q\,(t,\,x)\ =\ q_{\,\mathrm{f}}\,,
\end{equation*}
where $\D$ is the prescribed front celerity. Then, if we take $\D\ =\ \D_{\,\mathrm{num}}\,$, $\xi_{\,1}\ =\ u_{\,\mathrm{f}}\ +\ \sqrt{m_{\,\mathrm{f}}}\,$, it can be easily checked that the above solution corresponds fairly well to the unsteady solution shown in Figure~\ref{fig:3n}. In order to obtain the exact solution, we have to find the stationary solution to Equation~\eqref{eq:15}, \ie~the intersection point in the plane $(q,\,u)$ of two curves $\Gamma_{\,\Aa}$ and $\Gamma_{\,\Bb}$ defined as follows:
\begin{align*}
  u^{\,2}\ -\ m\ -\ (1\ +\ \delta)\,q^{\,2}\ &=\ 0\,, \qquad \bigl(\Gamma_{\,\Aa}\bigr) \\
  \alpha\,m\ -\ \sigma\,q\;\Bigl(u\ -\ \frac{m}{2\,(\D\ -\ u)}\Bigr)\ &=\ 0\,. \qquad \bigl(\Gamma_{\,\Bb}\bigr)
\end{align*}
Using the relation $m\ =\ \dfrac{\M}{\D\ -\ u}$ and the mass conservation condition at the wave front, we have
\begin{equation*}
  \M\ =\ (\D\ -\ u)\,m\ =\ \D\,m_{\,s}\,.
\end{equation*}
For parameter values used in our simulations, there exists a unique intersection point of curves $\Gamma_{\,\Aa}$ and $\Gamma_{\,\Bb}\,$, which satisfies the condition $(\D\ -\ u_{\,\mathrm{f}})^{\,2}\ >\ m_{\,\mathrm{f}}\,$. Such values can be easily found from System~\eqref{eq:tw1} -- \eqref{eq:tw4}.

In general, the questions of long time behaviour of unsteady solutions and the attraction property of similarity flows cannot be answered without a thourough additional scientific investigation. In particular, similar to the detonation theory \cite{Fickett1979}, it is necessary to formulate a criterium for the density current head front celerity selection. This celerity will determine the location of the wave front on large time intervals. In the following Sections we shall try to explain how to find the analytical approach to this problem.

%%% ----------------------------------------------------------------------- %%%

\section{Unsteady simulations of density currents with sediments entrainment}
\label{sec:unsteady}

In this Section we consider and analyze the results of numerical simulations of an underwater avalanche propagation along the slope. The process is sustained due to the sediments entrainment into the density flow. Namely, we solve numerically System \eqref{eq:11a} -- \eqref{eq:11d}. One of the questions we would like to study is the emergence of self-similar solutions from unsteady ones and dependence of the wave front celerity on initial perturbations. Moreover, in our numerical experiments we set the parameter $\kappa\ \equiv\ 0$ since this parameter does not change the qualitative behaviour of solutions. We notice also that parameter $\sigma$ determines only the ratio of vertical and horizontal flow scales. Thus, this parameter can be set to one by an appropriate choice of independent variables scaling. In this way we eliminate from our discussion unimportant parameters.

We assume that we have a constant slope ($\alpha\ =\ \const$), which is initially covered by a thin uniform and motionless layer of sediments with relative mass $m_{\,s}\ =\ b_{\,s}\,\zeta_{\,s}\ =\ \const$. On the left boundary of the channel a certain perturbation is prescribed. We assume that this perturbation is strong enough to initiate an underwater avalanche. In other words, the initially motionless sediments receive enough momentum during the passage of the avalanche front to be entrained into the density current. The sedimentation velocity can be neglected on time scales considered in our study.

In System \eqref{eq:11a} -- \eqref{eq:11d} we can turn to dimensionless variables by choosing the velocity scale as $\sqrt{m_{\,s}}$ along with the length scale $\ell\,$. We remark that the problem formulation does not contain any other dimensionless variables besides the bottom slope $\alpha\,$. Consequently, on a sufficient distance from the left boundary, we can assume that the flow will ``forget'' its initial conditions. In this case we may expect the emergence of self-similar solutions, whose asymptotic front celerity in dimensionless variables depends \emph{only} on the slope $\alpha\,$, \ie
\begin{equation}\label{eq:da}
  \D_{\,\mathrm{f}}\ =\ \Cc\,(\alpha)\,,
\end{equation}
where $\D_{\,\mathrm{f}}$ is the dimensionless celerity of the avalanche front ($m_{\,s}\ =\ 1$). This conjecture appears to be natural. However, our numerical simulations show that the situation is more complicated: the front velocity turns out to be dependent on other factors as well. Namely, we show that dependent on the initial generating conditions, the solution can induce \emph{at least two} different self-similar solutions with two different front velocities that we denote as
\begin{equation*}
  \Cc_{\,\ast}\,(\alpha)\ <\ \Cc^{\,\ast}\,(\alpha)\,.
\end{equation*}
This surprising property appears for a certain range of bottom slopes $\alpha\,$. In the present study we do not hope to investigate all possible situations leading to the hysteresis of self-similar solutions. Our goal is more modest being to study and to understand what we observe. Consequently, we focus on two particular values of the slope angle $\phi\ =\ 32^{\,\circ}$ and $\phi\ =\ 45^{\,\circ}\,$. This choice is not arbitrary. The mathematical model can be seriously validated for these two angles since the corresponding experiments on sediments entrainment by the avalanche front have been performed in \cite{Rastello2004}. The goal of our study is twofold. First of all, we would like to highlight the fact that the avalanche velocity is not unique. Then, we are going to explain the front velocity selection mechanism under hysteresis conditions by analysing the structure of travelling waves.

The numerical simulations are performed in the space-time domain $[\,0,\,L\,] \times [\,0,\,T\,]\,$. The initial conditions are chosen according to the classical lock-exchange problem:
\begin{equation*}
  u\,(0,\,x)\ =\ 0\,, \qquad q\,(0,\,x)\ =\ 0\,,
\end{equation*}
\begin{equation*}
  m\,(0,\,x)\ =\ \begin{dcases}
   \ m_{\,\ell}\,,& 0\ \leq\ x\ \leq\ \ell_{\,0}\,, \\
   \ 1,\,& \ell_{\,0}\ <\ x\ \leq\ L\,,
  \end{dcases}
  \qquad
  h\,(0,\,x)\ =\ \begin{dcases}
   \ h_{\,\ell}\,,& 0\ \leq\ x\ \leq\ \ell_{\,0}\,, \\
   \ h_{\,r},\,& \ell_{\,0}\ <\ x\ \leq\ L\,.
  \end{dcases}
\end{equation*}
The numerical parameters which are common to all simulations reported in this Section are given in Table~\ref{tab:params}. On left and right walls of the channel we impose simple impermeability conditions. The right boundary condition is not essential since the domain length $L$ is chosen such that the flow does not reach the right boundary during time $T\,$. In order to illustrate the hysteresis phenomenon, we report here two numerical simulations, which differ by the choice of the initial condition $h\,(x,\,0)\,$:
\begin{align*}
  (\Cc_{\,1})\,: & \qquad h_{\,\ell}\ =\ 1.05\,, \qquad h_{\,r}\ =\ 1.0\,, \\
  (\Cc_{\,2})\,: & \qquad h_{\,\ell}\ =\ 0.105\,, \qquad h_{\,r}\ =\ 0.1\,. \\
\end{align*}

\begin{figure}
  \centering
  \subfigure[$\phi\ =\ 32^{\circ}\,$, ($\Cc_{\,1}$)]{\includegraphics[width=0.48\textwidth]{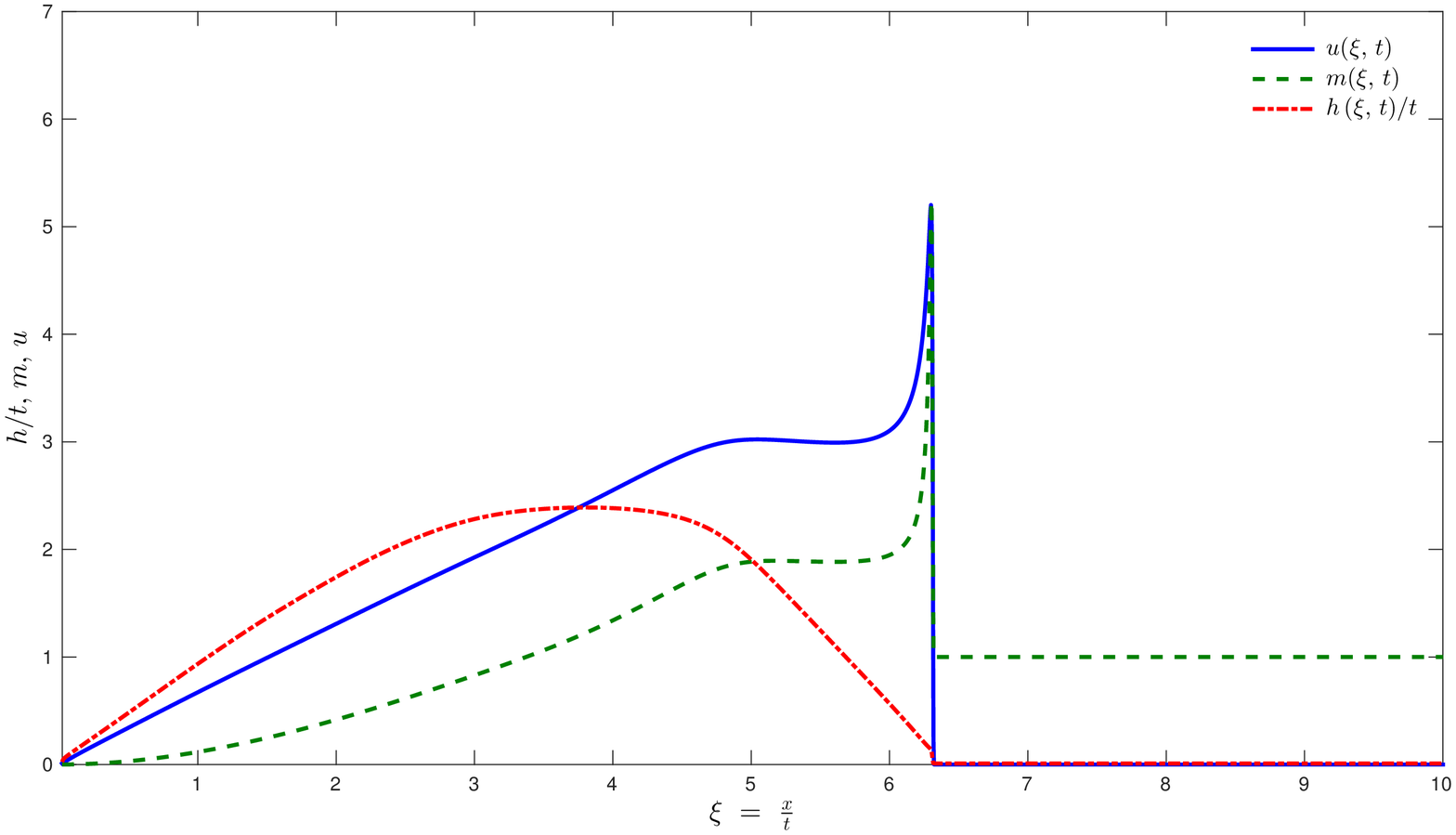}}
  \subfigure[$\phi\ =\ 32^{\circ}\,$, ($\Cc_{\,2}$)]{\includegraphics[width=0.48\textwidth]{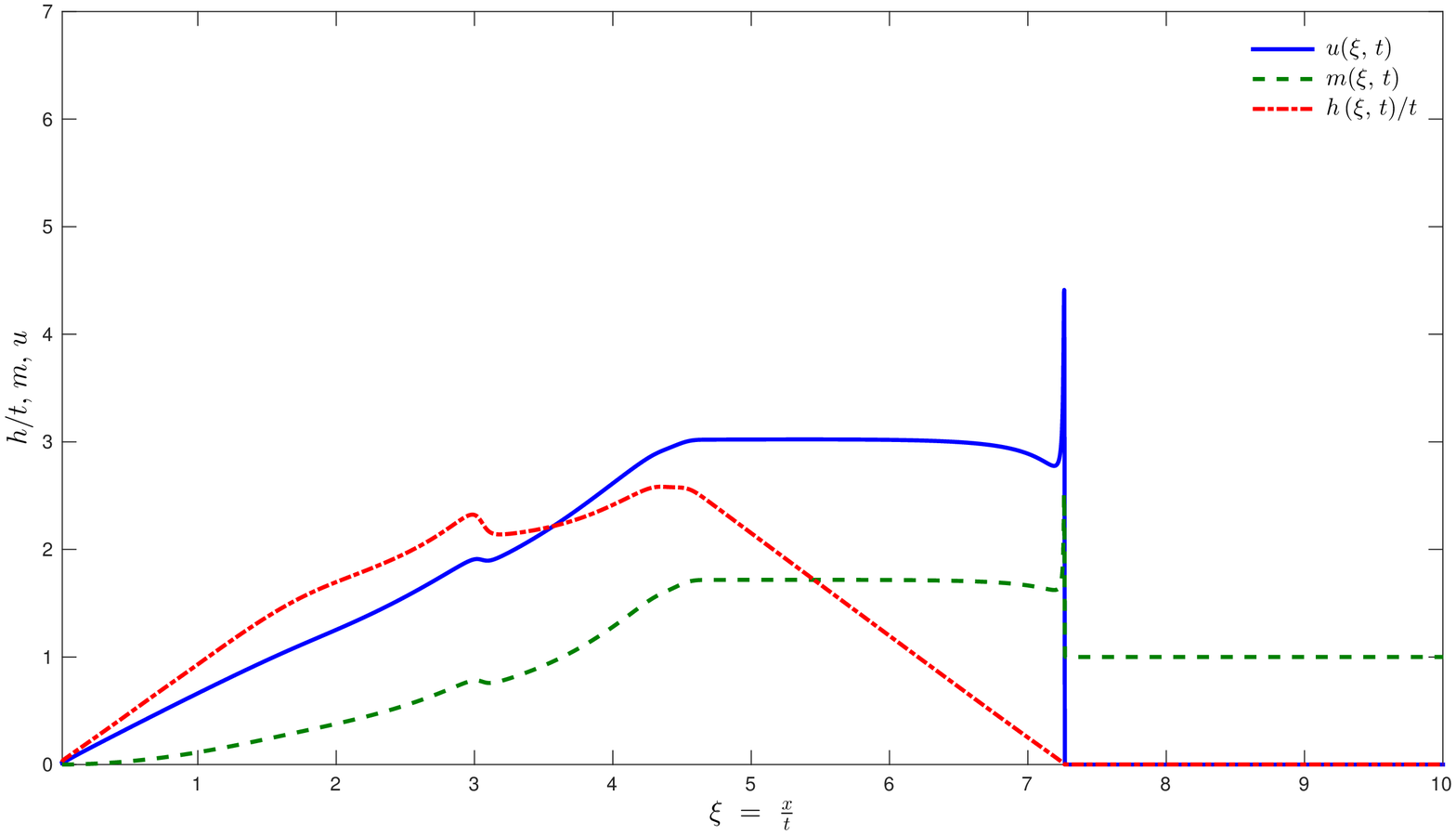}}
  \subfigure[$\phi\ =\ 45^{\circ}\,$, ($\Cc_{\,1}$)]{\includegraphics[width=0.48\textwidth]{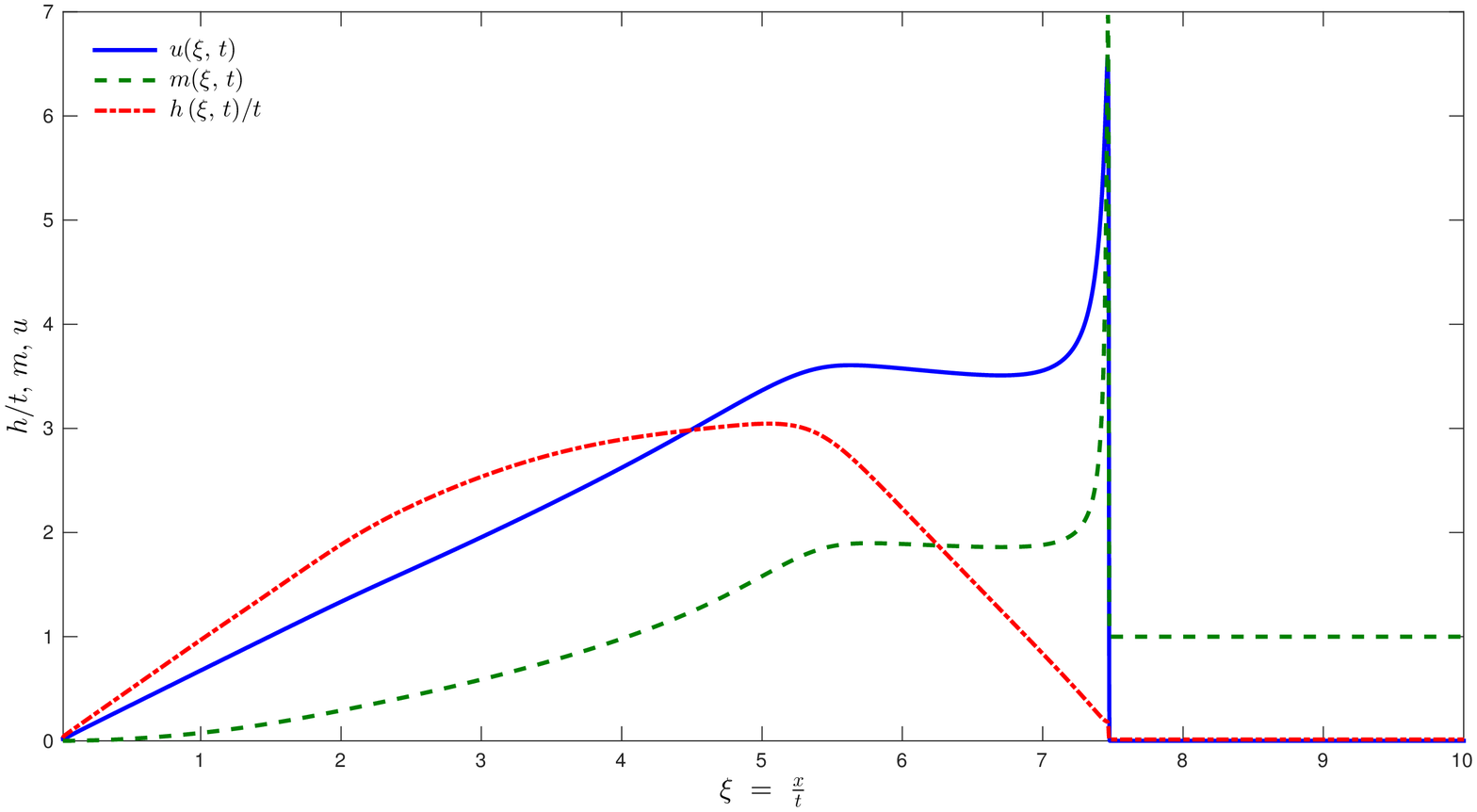}}
  \subfigure[$\phi\ =\ 45^{\circ}\,$, ($\Cc_{\,2}$)]{\includegraphics[width=0.48\textwidth]{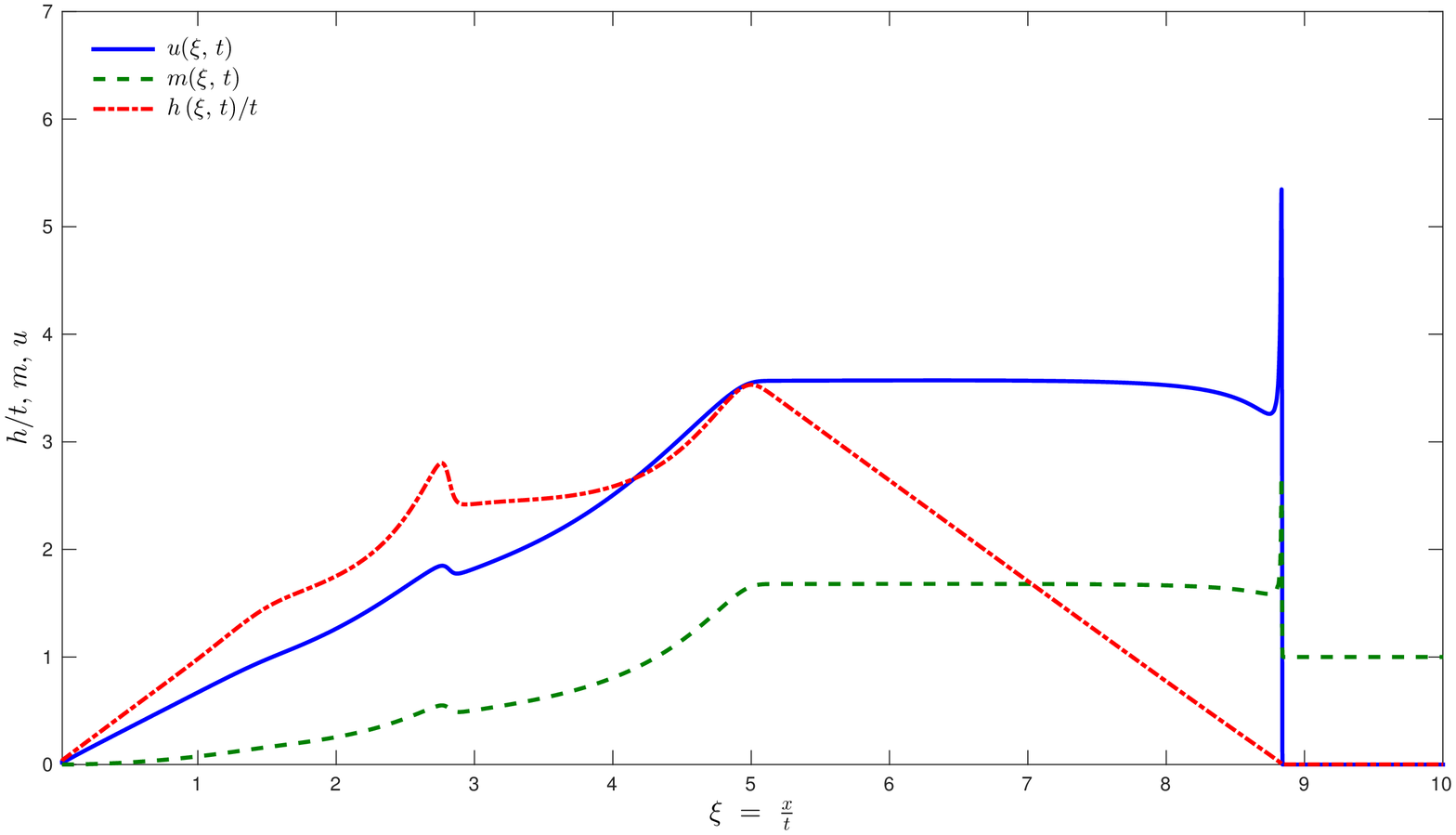}}
  \caption{\small\em Self-similar profiles emerging from the lock-exchange problem at the final simulation time $t\ =\ T\,$.}
  \label{fig:2}
\end{figure}

The results of these computations $(\Cc_{\,1,\,2})$ at final time $t\ =\ T$ are reported in Figure~\ref{fig:2} for two considered values of the angle $\phi\,$. All solutions are plotted in the appropriate self-similar variables $\bigl(t,\,\xi\ =\ \frac{x}{t}\bigr)$ in order to illustrate the emergence of self-similar solutions. We underline the fact that appropriate dependent variables are $\hat{h}\ \equiv\ h/t\,$, $u\,$, $m$ and $q$ (the mean turbulent velocity $q$ is not plotted in our Figures). We report also that for $t\ \gtrsim\ 0.1\cdot T\,$, the solutions practically do not depend anymore on $t\,$, when considered in self-similar variables.

\begin{table}
\begin{tabular}{l|c}
  \hline\hline
  \textit{Parameter} & \textit{Value} \\
  \hline\hline
  Computational domain length, $L$ & $12\,000$ \\
  Lock length, $\ell_{\,0}$ & $15$ \\
  Heavy fluid relative mass, $m_{\,\ell}$ & $1.05$ \\
  Final simulation time, $T$ & $1\,200$ \\
  \hline\hline
\end{tabular}
\bigskip
\caption{\small\em Numerical values of parameters used in lock-exchange flow simulations.}
\label{tab:params}
\end{table}

The further variation of initial conditions (in reasonable limits, of course) ($\Cc_{\,1,\,2}$) does not change the wave configurations together with observed front velocities. In this way, we demonstrate in the framework of System \eqref{eq:11a} -- \eqref{eq:11d} that numerical solutions possess two distinct asymptotic regimes whose front celerities are summarized in Table~\ref{tab:cast}. Obviously, the relative mass of sediments $m_{\,s}$ was kept the same in both experiments and it was distributed uniformly along the slope. The qualitative behaviour of solutions ($\Cc_{\,1}$) and ($\Cc_{\,2}$) is very similar. Right behind the front we can observe the `flat' region where the values of solutions are almost constant. The flow tail remains essentially unsteady. The main difference between solutions ($\Cc_{\,1}$) and ($\Cc_{\,2}$) consists in the fact that in the vicinity of the wave front the flow ($\Cc_{\,1}$) is transcritical, while the flow ($\Cc_{\,2}$) is supercritical in the frame of reference moving with the wave. This numerical observation is studied in more details in the following Section.

\begin{table}
\begin{tabular}{c|c|c}
\hline\hline
\multirow{2}{*}{\textit{Celerity}} & \multicolumn{2}{c}{\textit{Slope angle}} \\
\cline{2-3}
& $\phi\ =\ 32^{\,\circ}$ & $\phi\ =\ 45^{\,\circ}$ \\
\hline
$\Cc_{\,\ast}$ & $6.32$ & $7.4$ \\
$\Cc^{\,\ast}$ & $7.3$  & $8.9$ \\
\hline\hline
\end{tabular}
\bigskip
\caption{\small\em Numerically observed asymptotic front celerities in lock-exchange type experiments.}
\label{tab:cast}
\end{table}

%%% ----------------------------------------------------------------------- %%%

\section{Structure of travelling waves}
\label{sec:struct}

In this Section we investigate more the internal structure of travelling waves in the problem of density current propagation involving the entrainment of bottom sediments. The main problem consists in determining the flow head velocity $\D_{\,\mathrm{f}}$ by analysing solutions to System of ODEs \eqref{eq:tw1} -- \eqref{eq:tw4}, which describes all travelling wave solutions to System of PDEs \eqref{eq:11a} -- \eqref{eq:11d}.

Numerical simulations of unsteady solutions to Equations \eqref{eq:11a} -- \eqref{eq:11d} show the existence of some asymptotic regime provided that sediments are distributed uniformly along the slope, \ie~$m_{\,s}\ =\ \const$. In particular, Figure~\ref{fig:2} clearly demonstrates that the density current tends to a linear profile right behind the front:
\begin{equation}\label{eq:lin}
  h_{\,x}\ =\ \const, \qquad u\ =\ \const, \qquad m\ =\ \const, \qquad q\ =\ \const.
\end{equation}
This particular solution was analyzed in \cite{Liapidevskii2018}. We remark that the terminology might be slightly ambiguous since this linear solution belongs to the class of travelling waves \emph{and} to the class of similarity solutions with ansatz~\eqref{eq:ans}. In order to explain the velocity selection mechanism of the front celerity, we have to study in more details the solution right behind the wave front as shown in Figure~\ref{fig:2}. This solution in mathematical terms represents a travelling wave moving with the speed $\D_{\,\mathrm{f}}\,$. In the phase space this wave connects the state in front of the wave ($h\ =\ h_{\,r}\ \geq\ 0\,$, $m\ =\ m_{\,r}\ =\ 1\,$, $u\ =\ u_{\,r}\ =\ 0\,$, $q\ =\ q_{\,r}\ =\ 0$) with the linear solution \eqref{eq:lin}.

As it was demonstrated earlier using dimensional arguments, the problem of front celerity determination consists in finding the functional coefficient $\Cc\,(\alpha)\,$. In contrast to situations with constant mass inflow prescribed at the channel entrance (left boundary in our set-up), we do not possess enough experimental data for thermals to map the dependence \eqref{eq:da}. Consequently, we base our analysis on necessary conditions for the existence of travelling waves \eqref{eq:tw1} -- \eqref{eq:tw4} and their confrontation with unsteady simulations of governing Equations \eqref{eq:11a} -- \eqref{eq:11d}.

Consider a wave front, moving with constant \emph{supercritical} celerity $\D_{\,\mathrm{f}}\,$, which runs into quiescent constant density $\rho\ =\ \rho_{\,r}$ fluid layer:
\begin{equation*}
  m_{\,r}\ =\ \const\ >\ 0\,, \qquad
  h_{\,r}\ =\ \const\ \geq\ 0\,, \qquad
  u_{\,r}\ =\ 0\,, \qquad
  q_{\,r}\ =\ 0\,.
\end{equation*}
Thus, the condition that the celerity $\D_{\,\mathrm{f}}$ is supercritical can be written mathematically as
\begin{equation*}
  \D_{\,\mathrm{f}}\ >\ \sqrt{m_{\,r}}\ \equiv\ 1\,.
\end{equation*}
We have to analyze two possible situations in the following Sections.

%%% ----------------------------------------------------------------------- %%%

\subsection{Flow with a jump behind the head}
\label{sec:jump}

\begin{figure}
  \centering
  \subfigure[$(\Cc_{\,1})\,$, $\phi\ =\ 32^{\,\circ}\,$, $\D_{\,\mathrm{f}}\ \approx\ 6.32$]{\includegraphics[width=0.48\textwidth]{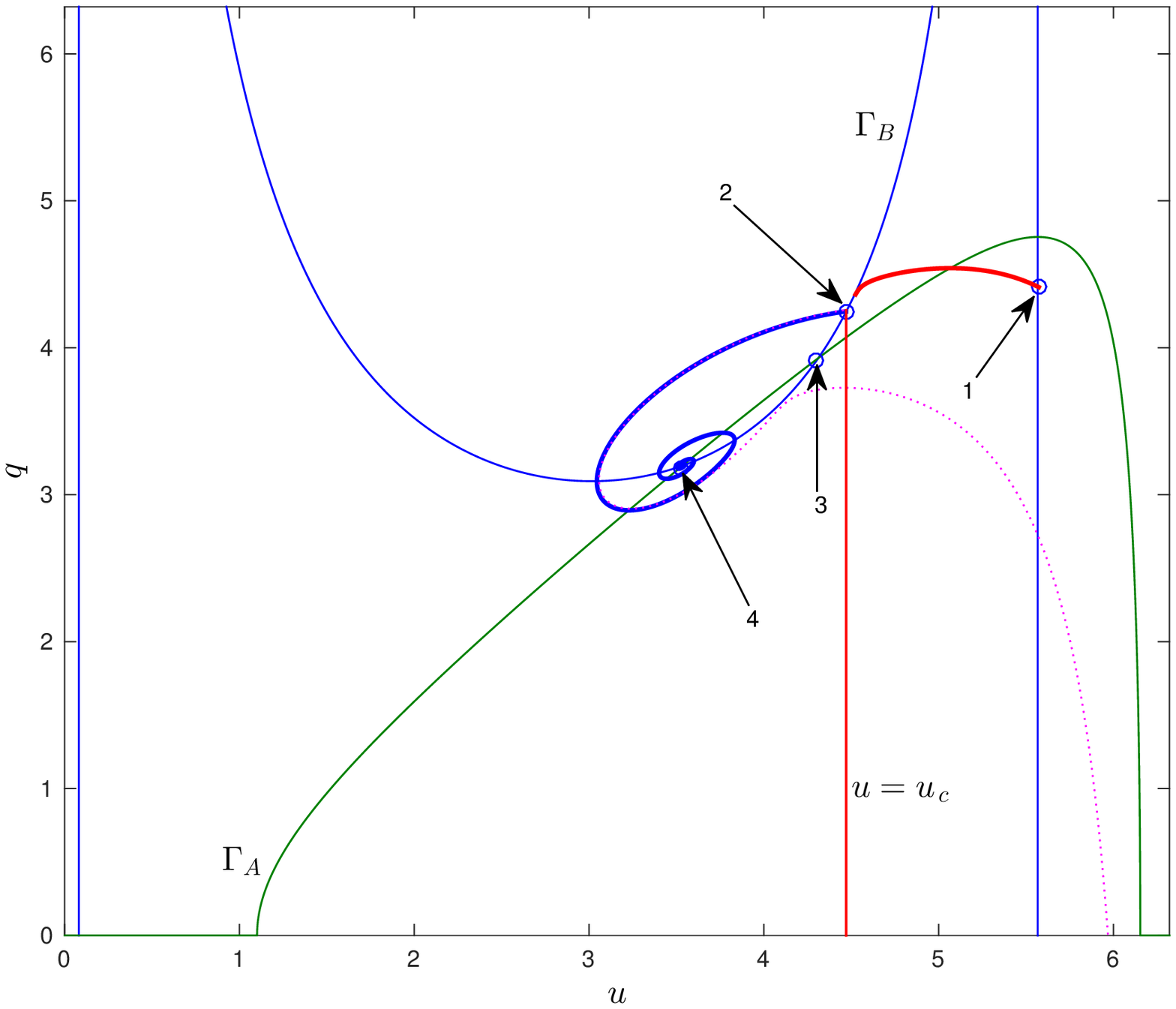}}
  \subfigure[$(\Cc_{\,2})\,$, $\phi\ =\ 32^{\,\circ}\,$, $\D_{\,\mathrm{f}}\ \approx\ 7.63$]{\includegraphics[width=0.48\textwidth]{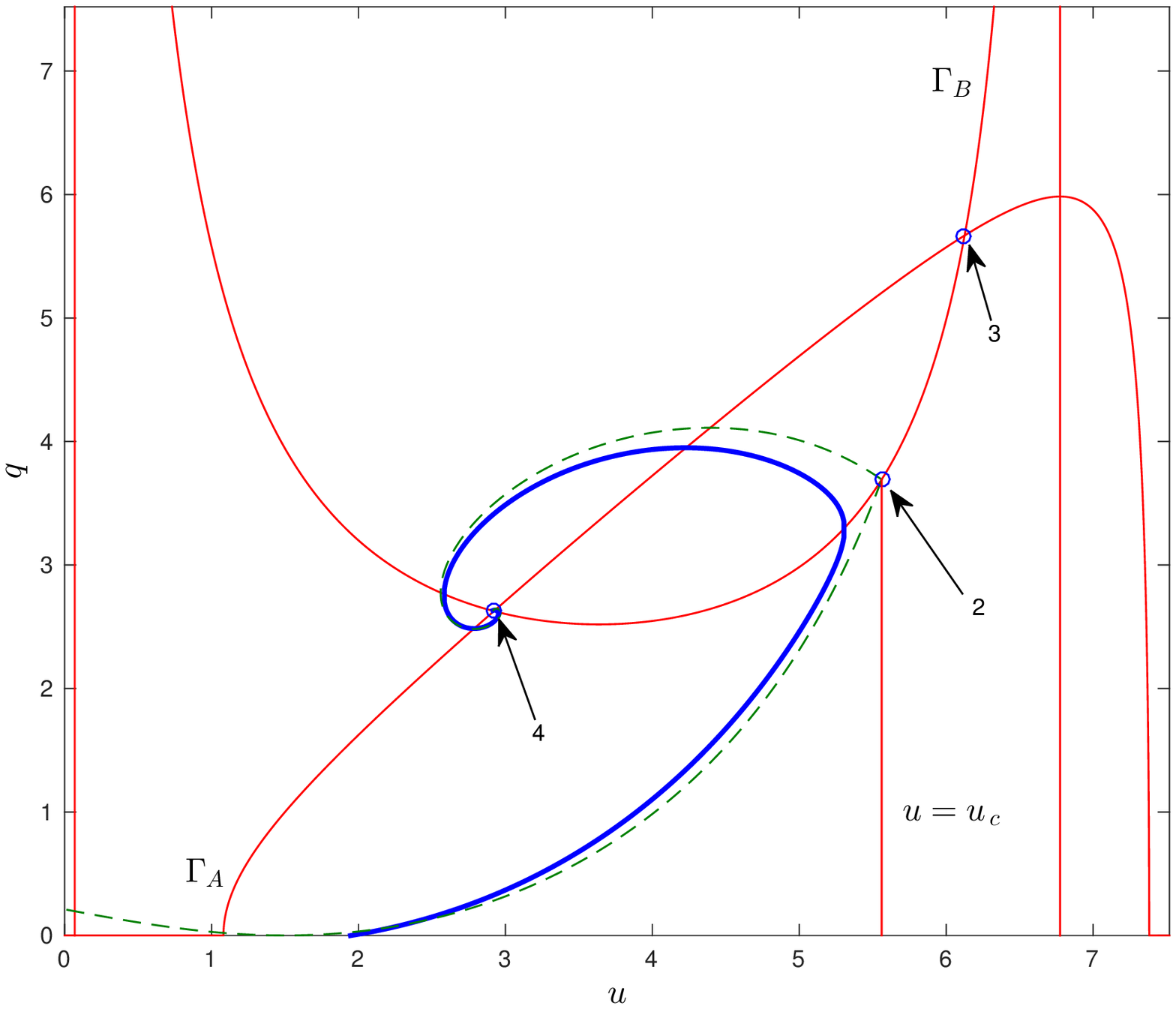}}
  \subfigure[$(\Cc_{\,1})\,$, $\phi\ =\ 45^{\,\circ}\,$, $\D_{\,\mathrm{f}}\ \approx\ 7.48$]{\includegraphics[width=0.48\textwidth]{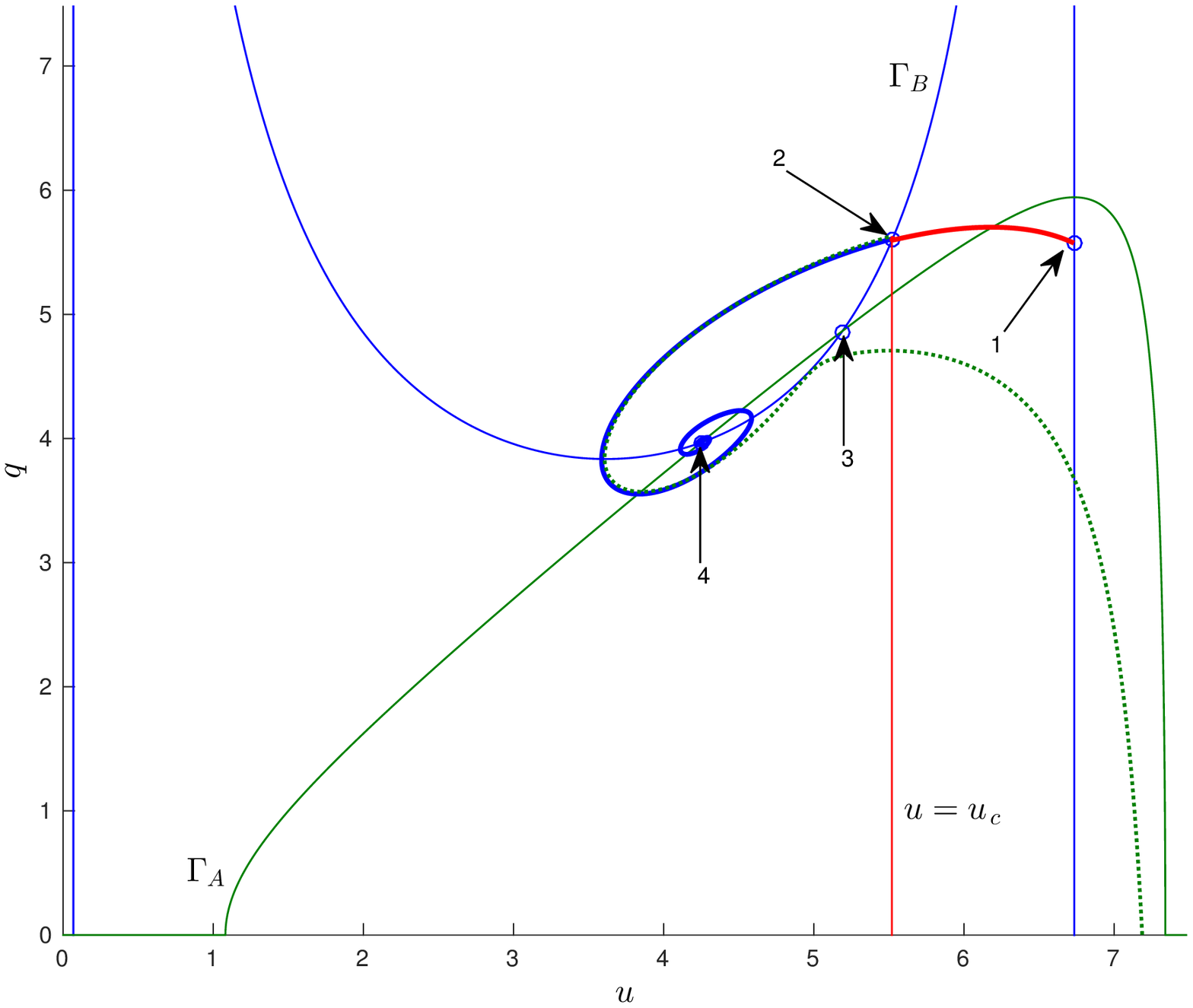}}
  \subfigure[$(\Cc_{\,2})\,$, $\phi\ =\ 45^{\,\circ}\,$, $\D_{\,\mathrm{f}}\ \approx\ 8.9$]{\includegraphics[width=0.48\textwidth]{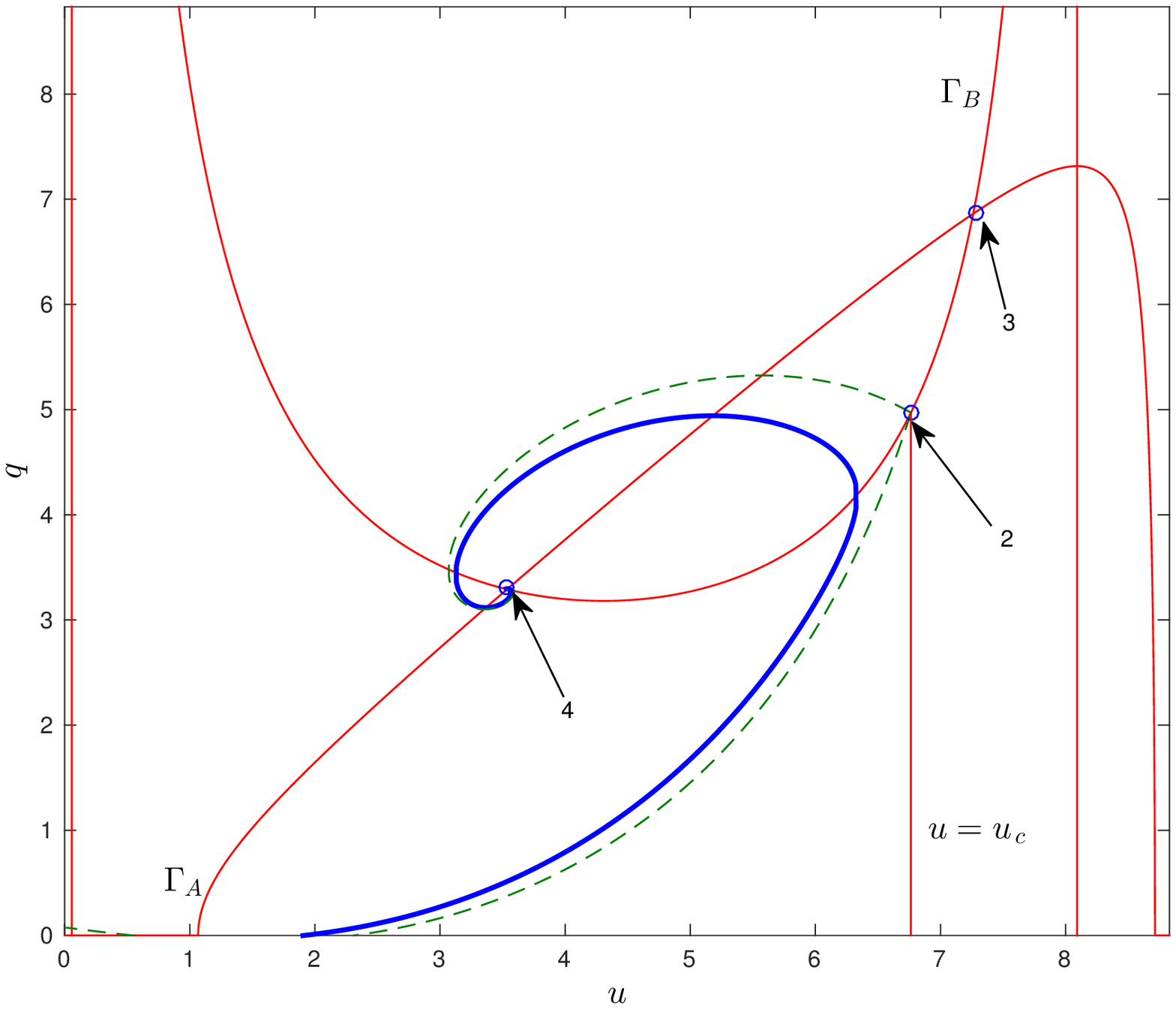}}
  \caption{\small\em Behaviour of thermals solutions in the phase plane $(u,\,q)$ for different slopes $\alpha\ =\ \tan\phi$ and different initial conditions $(\Cc_{\,1,\,2})\,$. Trans-critical are shown in panels (a) and (c), while super-critical flows in panels (b) and (d).}
  \label{fig:3}
\end{figure}

Behind the wave front the flow is sub-critical in the reference frame moving with the wave:
\begin{equation*}
  \D_{\,\mathrm{f}}\ -\ u_{\,1}\ <\ \sqrt{m_{\,1}}\,,
\end{equation*}
where by index ${}_{1}$ we denote the state right behind the wave front. This transition to the sub-critical flow behind the front implies the existence of a jump, which moves with the same velocity $\D_{\,\mathrm{f}}\,$. The state ${}_{1}$ can be determined from \textsc{Rankine}--\textsc{Hugoniot} conditions written for the System of balance laws \eqref{eq:11a} -- \eqref{eq:11d}:
\begin{align}\label{eq:24a}
  h_{\,1}\ &=\ \frac{1}{2}\;\bigl(\sqrt{h_{\,r}^{\,2}\ +\ 8\,\D_{\,\mathrm{f}}^{\,2}\,\frac{h_{\,r}}{b_{\,r}}}\ -\ h_{\,r}\bigr)\,, \\
  u_{\,1}\ &=\ \frac{\D_{\,\mathrm{f}}\,(h_{\,1}\ -\ h_{\,r})}{h_{\,1}}\,, \label{eq:24b} \\
  m_{\,1}\ &=\ b_{\,r}\,h_{\,1}\,, \label{eq:24c} \\
  q_{\,1}\ &=\ \sqrt{\D_{\,\mathrm{f}}^{\,2}\ +\ 2\,b_{\,r}\,h_{\,r}\ -\ u_{\,1}^{\,2}\ -\ 2\,b_{\,r}\,h_{\,1}}\,.\label{eq:24d}
\end{align}
In Figure~\ref{fig:3} we show the structure of the flow behind the wave front in the phase plane $(u,\,q)\,$. The graphs (\textit{a,\,b}) represent the case $\phi\ =\ 32^{\,\circ}\,$, while panels (\textit{c,\,d}) correspond to the case $\alpha\ =\ 1\ \Longleftrightarrow\ \phi\ =\ 45^{\,\circ}\,$.

Let us explain notation from Figure~\ref{fig:3}. Point $(1)$ corresponds to the state behind the jump, which is determined by formulae \eqref{eq:24a} -- \eqref{eq:24d}. Point $(2)$ is the intersection of the curves $\Gamma_{\,B}$ and $\Gamma_{\,C}\,$. We remind that vertical line $\Gamma_{\,C}$ separates the region of super-critical ($u\ <\ u_{\,c}$) from sub-critical ($u\ >\ u_{\,c}$) flows. Finally, points $(3)$ and $(4)$ are intersections of curves $\Gamma_{\,A}$ and $\Gamma_{\,B}\,$. The last two curves $\Gamma_{\,A,\,B}$ are always represented by solid lines in Figure~\ref{fig:3}. We underline that points $(2)\,$, $(3)\,$ and $(4)$ are critical points for the dynamical system \eqref{eq:15}. Consider the case of trans-critical flow ($\Cc_{\,1}$) in more details. A continuous sub-critical solution departing from point $(1)$ arrives to the critical point $(2)\,$. Super-critical solution ($u\ <\ u_{\,c}$) departing from $(2)$ arrives to the critical point $(4)\,$, as it is shown in Figures~\ref{fig:3}(\textit{a}) and \ref{fig:3}(\textit{c}). Thus, at certain distance from the wave front a solution to ODE system \eqref{eq:tw1} -- \eqref{eq:tw4} takes a very simple form:
\begin{equation}\label{eq:25}
  u\ =\ u_{\,4}\,, \qquad
  m\ =\ m_{\,4}\,, \qquad
  q\ =\ q_{\,4}\,, \qquad
  \lambda\ \eqdef\ \od{h}{x}\ =\ \const\ <\ 0\,.
\end{equation}
The flow, which is determined by last relations, is super-critical in the frame of reference moving with the speed $\D_{\,\mathrm{f}}\,$, \ie~$u_{\,4}\ <\ u_{\,c}\,$. In general, for a given slope $\alpha\,$, this solution depends also on the celerity $\D_{\,\mathrm{f}}\,$. We remark also that in view of formula~\eqref{eq:25}, this solution belongs also to the class of self-similar solutions with $\vtheta\ =\ 0$ (see Section~\ref{sec:ans}).

The choice of initial conditions $m\ =\ m_{\,s}$ and $h\ =\ h_{\,s}$ (we remind that $u_{\,s}\ =\ 0$ and $q_{\,s}\ =\ 0$) does not influence the resulting unsteady solutions since we can always choose dimensionless variables so that $m_{\,s}\ =\ 1$ and $h_{\,s}\ =\ 1\,$. In the present study we shall not perform an exhaustive analysis of all possible configurations of critical points and integral curves departing from point $(1)$ of System \eqref{eq:tw1} -- \eqref{eq:tw4} with various values of $\D_{\,\mathrm{f}}\,$. We underline also the fact that considered above solutions exist only for a certain interval of speeds:
\begin{equation*}
  \D_{\,\ast}\ <\ \D_{\,\mathrm{f}}\ <\ \D_{\,1}\,.
\end{equation*}
We are interested in determining the lower bound $\D_{\,\ast}$ for which a solution passing through the point $(1)$ arrives to the point $(4)$ (see again Figures~\ref{fig:3}(\textit{a}) and \ref{fig:3}(\textit{c})). Such value $\D_{\,\ast}$ is estimated from direct numerical simulations by making a standard dichotomy search in parameter $\D_{\,\mathrm{f}}\,$. In Figures~\ref{fig:3}(\textit{a}) and \ref{fig:3}(\textit{c}) with solid lines we show the solutions corresponding to limiting values of $\D_{\,\mathrm{f}}\ =\ \D_{\,\ast}\,$: $\D_{\,\ast}\ =\ 6.32$ for $\phi\ =\ 32^{\,\circ}$ and $\D_{\,\ast}\ =\ 7.48$ for $\phi\ =\ 45^{\,\circ}\,$. The dotted lines show the qualitative behaviour of solutions to ODE \eqref{eq:15} with $\D_{\,\mathrm{f}}\ <\ \D_{\,\ast}\,$. Such solutions cannot be realized as the solution of the full System \eqref{eq:tw1} -- \eqref{eq:tw4}, since the determinant of the system vanishes at $u\ =\ u_{\,c}\,$. It was shown by numerical calculations in Section~\ref{sec:valid} that the avalanche head tends asymptotically in time to the self-similar solution described by parameter $\vtheta\ =\ 0$ and $\D_{\,\mathrm{f}}\ =\ \D_{\,\ast}\,$. Thus, in dimensional variables we can state
\begin{equation*}
  \D_{\,\mathrm{f}}\ =\ \Cc_{\,\ast}\,(\alpha)\cdot \sqrt{m_{\,s}}\,,
\end{equation*}
with $\Cc_{\,\ast}\ \approx\ 6.32$ for $\phi\ =\ 32^{\,\circ}\,$, $\Cc_{\,\ast}\ \approx\ 7.48$ for $\phi\ =\ 45^{\,\circ}\,$. It means that for the initial data considered in Section~\ref{sec:valid}, the case of the trans-critical flow behind the front ($\Cc_{\,1}$) is realized (see the estimations provided in Equation~\eqref{eq:31}).

%%% ----------------------------------------------------------------------- %%%

\subsection{Continuous self-sustaining flow}

If we increase the parameter $\D_{\,\mathrm{f}}\ >\ \D_{\,\ast}\,$, another configuration of critical points in the phase plane $(u,\,q)$ is realized. The line $u\ =\ u_{\,c}$ corresponds to the transition of the solution through a critical state. This line also separates points $(3)$ and $(4)$. Moreover, a continuous and super-critical\footnote{As always, we refer to the super- and sub- criticality properties in the reference frame moving with the wave front.} solution departing from point $(0,\,0)$ always reaches the critical point $(4)$. In Figures~\ref{fig:3}(\textit{b}) and \ref{fig:3}(\textit{d}) it is illustrated with thick solid lines. In this way, the unsteady solution to the considered problem also tends asymptotically in time to the self-similar state \eqref{eq:25}. However, there is one important difference: the wave front celerity $\D_{\,\mathrm{f}}$ is substantially faster than for solutions considered in the previous Section~\ref{sec:jump}.

In order to determine the minimal value of the front celerity $\D^{\,\ast}$ for which a continuous solution to our problem is still realized, we observe that when we decrease the value of $\D_{\,\mathrm{f}}\,$, the continuous super-critical solution emanating from $(0,\,0)$ approaches a singular curve connecting points $(2)$ and $(4)$ (represented with a dashed line in Figures~\ref{fig:3}(\textit{b}) and \ref{fig:3}(\textit{d})). Further decrease of the front celerity $\D_{\,\mathrm{f}}$ results in the intersect of the solution to ODE \eqref{eq:15} and the critical line $\Gamma_{\,C}\,$. Therefore, such a solution cannot represent a bounded solution to System \eqref{eq:tw1} -- \eqref{eq:tw4} (see the previous Section~\ref{sec:jump}). Consequently, the minimal value $\D_{\,\mathrm{f}}\ =\ \D^{\,\ast}$ is achieved when the continuous solution coincides with this singular curve. If we come back to dimensional variables, we obtain the following dependence:
\begin{equation*}
  \D_{\,\mathrm{f}}\ =\ \Cc^{\,\ast}\,(\alpha)\cdot \sqrt{m_{\,s}}\,,
\end{equation*}
with $\Cc^{\,\ast}\,(\alpha)\ >\ \Cc_{\,\ast}\,(\alpha)\,$, $\Cc^{\,\ast}\ \approx\ 7.63$ for $\phi\ =\ 32^{\,\circ}$ and $\Cc^{\,\ast}\ \approx\ 8.9$ for $\phi\ =\ 45^{\,\circ}\,$. The detailed analysis of the dependences $\Cc^{\,\ast}\,(\alpha)$ and $\Cc_{\,\ast}\,(\alpha)$ will be a topic of our future investigations.

We performed the analysis of stationary and unsteady solutions, describing the evolution and propagation of density gravity flows along moderate slopes. This process is sustained due to the entrainment of sediments. The model was validated for two different values of the bottom slope. These values were chosen to match with available experimental data \cite{Rastello2004}. Nevertheless, the obtained results allow us to formulate the following rule for the flow head velocity:
\begin{quote}
\textit{For every regime of the flow, such as continuous super-critical or discontinuous trans-critical flow (in the frame of reference moving with the wave) with a jump at the front, the minimal possible celerity is realized in practice, provided that with this celerity we remain in the given class of solutions.}
\end{quote}
At the current stage, the statement given above should be considered as a \emph{conjecture} supported by strong numerical evidences along with the phase plane analysis. Moreover, this conjecture seems to be very plausible, since it reminds another well-established physical rule --- the least action principle \cite{Landau1976}. In future studies we are going to work on further validations of this rule. In particular, we would like to understand the limits of its applicability. Finally, we would like to make a couple of remarks.

\begin{remark}
The analysis of travelling waves presented hereinabove along with the front celerity selection rule was presented on the phase plane $(u,\,q)\,$. That is why the curves shown in Figure~\ref{fig:3} should be considered as projections of such solutions on the plane $(u,\,q)\,$. Nevertheless, these curves provide us with some information about solution properties connecting the starting point $\bigl(h_{\,r},\,m_{\,r},\,0,\,0\bigr)$ with terminal point $\bigl(\hat{h}_{\,4},\,m_{\,4},\,u_{\,4},\,q_{\,4}\bigr)$ only in the case where such solutions exist. Consequently, the considered proportionality coefficients $\Cc_{\,\ast}\,(\alpha)$ and $\Cc^{\,\ast}\,(\alpha)$ make sense only under this existence condition. Nevertheless, the obtained criterium of the celerity selection is fully confirmed by our numerical simulations and by extensive comparisons of the structure of travelling waves with obtained unsteady solutions.
\end{remark}

\begin{remark}
The procedure to determine critical velocities in the plane $(u,\,q)$ can be simplified if we seek the required value of the parameter $\D_{\,\mathrm{f}}$ from the condition:
\begin{description}
  \item[$\D_{\,\mathrm{f}}\ =\ \D_{\,\ast}\;$] the singular curve passes through the critical point $(1)\,$,
  \item[$\D_{\,\mathrm{f}}\ =\ \D^{\,\ast}\;$] the singular curve passes through the point $(0,\,0)\,$.
\end{description}
\end{remark}

%%% ----------------------------------------------------------------------- %%%

\section{Discussion}
\label{sec:disc}

Above we presented our developments on the modelling of shallow underwater turbidity currents. Below we outline the main conclusions and perspectives of our study.

%%% ----------------------------------------------------------------------- %%%

\subsection{Conclusions}

In the present manuscript we studied a long wave model including entrainment effects of the ambient liquid and of bottom sediments. The structure of self-similar and travelling wave solutions to the turbidity currents model in an inclined channel was investigated as well. This model was proposed earlier in \cite{Liapidevskii2004, Liapidevskii2018}. In particular, the focus here was on the determination of the front velocity problem, which was demonstrated to lack the important property of the unicity. The main peculiarity of this model consists in the fact that the entrainment rate can be determined from the governing equations to produce a closed and self-consistent set of equations. For this system there is a wide class of special exact solutions (stationary, travelling and self-similar ones). These solutions provide the asymptotic behaviour of turbidity currents involving sediments entrainment for large evolution times. For example, we demonstrated that the predicted self-similar shapes emerge naturally from quite generic unsteady solutions. The comparison with experimental data from \cite{Rastello2004} showed that the model describes correctly both the acceleration and deceleration stages of thermic propagation over fairly moderate slopes (\ie~$\phi\ \propto\ 30^{\,\circ}\ \sim\ 45^{\,\circ}$). However, the hysteresis phenomenon was shown to be present and it complicates significantly the prediction of the avalanche head velocity. To our knowledge, this is the first study which highlights this phenomenon. Nevertheless, for each flow regime (sub- or super-critical) we formulated the rules of the front velocity, which is the main parameter we need to determine for gravity-driven underwater avalanches. In particular, the phase plane $(u,\,q)$ analysis showed that two types of situations are possible (depending on the front speed): (\textit{i}) continuous flows and (\textit{ii}) flows with a discontinuity at the front. Every type of solutions can be realized as a large time asymptotics of an unsteady numerical simulation of a lock-exchange problem. Moreover, the minimal admissible front velocity is attained in numerical simulations, which allowed us to formulate the velocity selection rules similar to the \textsc{Chapman}--\textsc{Jouguet} principle in the detonation theory \cite{Chapman1899}.

We would like to mention also that far from the heavy fluid release area there is only one parameter having the dimension of the velocity --- $\sqrt{m_{\,s}}\,$. Consequently, the steady front speed should be proportional to this quantity, \ie~$\D_{\,\mathrm{f}}\ =\ c\,\sqrt{m_{\,s}}\,$. However, for different types of solutions, the proportionality coefficient $c$ is different. Moreover, it depends also on the initial disturbance, which initiated the flow. This dependence was referred to in our study as the hysteresis phenomenon.

%%% ----------------------------------------------------------------------- %%%

\subsection{Perspectives}

In the present manuscript we considered the two-dimensional ($2$D) physical problem, which was modelled by a one-dimensional hydrostatic ($1$DH) mathematical model. In our future works we would like to extend the present methods to $3$D physical situations which will be modelled by $2$DH models. Moreover, we assumed in the present study that the ambient fluid layer has infinite depth for the sake of simplicity. In future studies we would like to investigate the finite depth effects of the upper layer on the avalanche shape and velocity. We know already that decreasing the slope angle increases the influence of finite channel height effects on the front velocity. This study will require the explicit introduction of the second layer into our model. Decreasing the slope angle has another potential difficulty: the growth of the flow head via the ambient fluid entrainment into it may cause the local loss of model hyperbolicity and, thus, the development of flow instability. So, these issues need to be faced in future studies.

Another natural extension of the present work is to consider uneven bathymetries since underwater obstacles will perturb the avalanche propagation and the interaction between these two entities has to be studied. Finally, the inclusion of non-hydrostatic effects is also desirable to improve the accuracy of our physical model.

%%% ----------------------------------------------------------------------- %%%

\subsection*{Acknowledgments}
\addcontentsline{toc}{subsection}{Acknowledgments}

This research was supported by CNRS under the project PEPS \'Energie 2017 (project MN4BAT), the Russian Science Foundation (project 15--11--
20013) along with the Interdisciplinary Program II.1 of SB RAS (project no.~2). V.~\textsc{Liapidevskii} acknowledges the hospitality of the University \textsc{Savoie Mont Blanc} during his visit in December 2017.

%%% ----------------------------------------------------------------------- %%%

\appendix
\section{Mathematical model derivation}
\label{app:der}

Consider an incompressible liquid which fills a two-dimensional fluid domain $\Omega$. A \textsc{Cartesian} coordinate system $O\,\x$, $\x\ =\ (x,\,y)$ is chosen in a classical way such that the axis $Oy$ points vertically upward and abscissa $O\,x$ is positive along the right horizontal direction. The fluid is bounded below by a solid non-erodible bottom $y\ =\ d\,(x)\,$. Above, the fluid can be assumed unbounded for the sake of simplicity, since our attention will be focused on processes taking place in the region close to the bottom.

The fluid is inhomogeneous and the flow can be conventionally divided in three parts. On the solid bottom there is a heavy fluid layer with density $\rho_{\,0}$ composed mainly of sedimentary deposits. Its thickness is $\zeta\,(x,\,t)\,$. Above, we have a muddle layer whose density will be denoted by $\rhob\,(\x,\,t)$ composed of the sediments and the still water mix. Its thickness is $h\,(x,\,t)\,$. Finally, the whole domain above these two layers is filled with the still ambient water of the density $\rho_{\,a}\ >\ 0\,$. We will initially assume that the ambient fluid is at rest ($\u_{\,a}\ \equiv\ \vO$). This assumption will be relaxed later in Section~\ref{sec:amb}. This situation is schematically depicted in Figure~\ref{fig:sketch}. We include also into consideration the situation where a thin motionless layer of sediments with constant thickness $h_{\,s}$ and density $\rho_{\,s}\ >\ 0$ covers the slope\footnote{If the layer of sediments vanishes before the avalanche front, the flow will degenerate into exchange or thermal flow depending on the boundary conditions specified at the entrance of the computational domain. These cases were considered in our previous publication \cite{Liapidevskii2018}.}. This assumption is needed to initiate and study self-sustained flows down the slope. It is usually the case in many practical situations and these sediment deposits may contribute to the flow head dynamics while propagating along the incline. One may imagine also that a certain mass of a sediment suspension or dense fluid is released into the flow domain at $x\ =\ 0$ with the mass flux $\rho_{\,0}\cdot\zeta\,(0,\,t)\cdot u\,(0,\,t)\,$.

\begin{remark}
L.~\textsc{Ovsyannikov} (1979) showed \cite{Ovsyannikov1979} that the quiescent still water layer can be indifferently chosen to have a free surface or to be bounded above by a rigid wall. Some authors assume the latter case \cite{DeLuna2008, MoralesdeLuna2009}. Both boundary conditions lead to the same mathematical long wave model under the \textsc{Boussinesq} assumption\footnote{In turbidity density currents the particle concentrations are in general sufficiently low (with $0.1$ -- $7\%$ of the volume \cite{Meiburg2010}). In such concentrations particle/particle interactions are negligible \cite{Bagnold1954}. Thus, from the modelling point of view the \textsc{Boussinesq} approximation is generally adopted.} which will be adopted below. So, in the present study for the sake of convenience we choose the infinite still water layer.
\end{remark}

If we assume the fluid to be perfect, its flow can be described by classical incompressible \textsc{Euler} equations. The variable fluid density $\rho(\x,\,t)$ and the fluid velocity field $\u(\x,t)\ \eqdef\ \bigl(u(\x,t),\, v(\x,t)\bigr)$ satisfy the following system of equations:
\begin{eqnarray*}
  \div\u\ &=&\ 0\,, \\
  \rho_t\ +\ \div[\,\rho\,\u\,]\ &=&\ 0\,, \\
  (\rho\,\u)_t\ +\ \div[\,\rho\,\u\otimes\u\ +\ p\,\Id\,]\ &=&\ \rho\,\g\,,
\end{eqnarray*}
where the subscript $t$ denotes the differentiation with respect to time and the operator $\grad\ \eqdef\ (\partial_x,\, \partial_y)$ is the gradient, $p\,(\x,t)$ is the fluid pressure variable and $\Id\ =\ (\delta_{ij})_{1\,\leq\,i,\,j\,\leq\,2}$ is the identity tensor compactly written with the \textsc{Kronecker} $\delta$-operator. Taking into account the choice of the coordinate system, the gravity acceleration vector is directed vertically downwards, \ie~$\g\ =\ (0,\, -\,g)\,$. The ideal fluid assumption can be also seen as a flow with an infinitely large \textsc{Reynolds} number $\RE\ \to\ +\infty$ \cite{Lamb1932, Stoker1958}. So, the flow is a priori turbulent and we shall return to this question below.

In gravity flows the particle\footnote{Here we speak about heavy `particles', which constitute the sediment layer.} density is generally larger, but the resulting muddy mixture turns out to be of the same order of magnitude with the ambient fluid. Thus, in turbidity flows modelling it is common to assume that the density variations inside the fluid column are not large, \ie
\begin{equation}\label{eq:bouss}
  \frac{\rho_{\,0}\ -\ \rho_{\,a}}{\rho_{\,a}}\ \ll\ 1\,.
\end{equation}
In these conditions the so-called \textsc{Boussinesq} approximation can be applied \cite{Fukushima1990, Naaim-Bouvet2003}, which consists in taking into account the density variations only in the buoyancy term. This approximation results in the following system of equations:
\begin{eqnarray}\label{eq:bouss1}
  \div\u\ &=&\ 0\,, \\
  b_{\,t}\ +\ \div[\,b\,\ut\,]\ &=&\ 0\,, \\
  \u_{\,t}\ +\ \div\bigl[\,\u\otimes\u\ +\ \frac{p}{\rho_{\,a}}\;\Id\,\bigr]\ &=&\ \b\,, \label{eq:bouss2}
\end{eqnarray}
where $\ut$ is the buoyancy transport velocity and, for the sake of convenience, we replace the density variations $\dfrac{\rho\ -\ \rho_{\,a}}{\rho_{\,a}}$ by the so-called buoyancy function $b\,$:
\begin{equation}\label{eq:buo}
  b\,(\x,\,t)\ \eqdef\ \frac{\rho\ -\ \rho_{\,a}}{\rho_{\,a}}\; g\; \pi\,(\x,\,t)\,, \qquad \b\,(\x,\,t)\ \eqdef\ \bigl(0,\, -\,b\,(\x,\,t)\bigr)\,.
\end{equation}
We note that the variable density in the flow can be realized by the variable concentration of sediments or by dense fluid input. The function $\pi\,(\x,\, t)$ is the concentration of the sediments in the suspension \cite{Ishii2006, Dias2008, Meyapin2009}. For instance, the concentration $\pi\,(\x,\,t)$ is equal identically to zero in the still pure water region and it is equal to one in the bottom sediments layer. However, sometimes the sediments porous nature needs to be taken into account \cite{Meyer-Peter1948, Einstein1952, Grass1981, Richardson1985, Nielsen1992}. In our modelling paradigm it can be achieved through choosing the concentration value $\pi\,(\x,\,t)\ \equiv\ \pi_{\,0}\ \apprle\ 1$ in the sedimentary layer slightly smaller than one.

Finally, the buoyancy transport velocity $\ut$ incorporates the sediment fall velocity $v_{\,s}\ \geq\ 0\,$, which is assumed to be constant in our study:
\begin{equation*}
  \ut\ \eqdef\ (u,\ v\ -\ v_{\,s})\,.
\end{equation*}
Some empirical consideration about the determination of the parameter $v_{\,s}$ can be found in \cite{Dietrich1982}, for example. System \eqref{eq:bouss1} -- \eqref{eq:bouss2} possesses an energy conservation law which can be readily obtained:
\begin{equation}\label{eq:energy}
  \bigl(\half\,\abs{\u}^2\ +\ b\,y\bigr)_{\,t}\ +\ 
  \div\Bigl[\,\bigl(\half\,\abs{\u}^2 + b\,y\bigr)\,\u\ +\ \frac{p\,\u}{\rho_a}\,\Bigr]\ =\ y\,(b\,v_s)_{\,y}\,,
\end{equation}
where $\abs{\u}^{\,2}\ \equiv\ u^{\,2}\ +\ v^{\,2}$ denotes the usual \textsc{Euclidean} distance. The subscript $(\cdot)_{\,y}$ in the right hand side denotes the partial derivative with respect to $y\,$. We underline the fact that Equation \eqref{eq:energy} is a direct differential consequence of the governing Equations \eqref{eq:bouss1} -- \eqref{eq:bouss2}. So, up to this point Equation \eqref{eq:energy} is redundant.

\begin{remark}
Earlier we introduced the buoyancy $b\,(\x,\,t)$ through the variable concentration of sediments. However, this quantity can be introduced independently of the physical mechanism, which creates the density gradient. For instance, the density current can be created by injecting a heavy fluid into the light ambient. In this case the quantity $\pi\,(\x,\,t)$ should be understood as the heavy fluid volumetric concentration.
\end{remark}

\begin{remark}
In the description of density currents some other quantities can be introduced. In our study we chose the buoyancy to describe density variations. However, other authors may opt for other choices. The goal of this Remark is to introduce some alternatives to the reader. For instance, one can consider the \emph{effective} gravitational acceleration \cite{Turner1973}:
\begin{equation*}
  \tilde{g}\ \eqdef\ g\;\frac{\rho_{\,0}\ -\ \rho_{\,a}}{\rho_{\,a}}\ \equiv\ g\;\frac{\Delta\,\rho}{\rho_{\,a}}\,.
\end{equation*}
Then, the densimetric \textsc{Froude} number can be defined:
\begin{equation*}
  \Fr_{\,\mathrm{d}}\ \eqdef\ \frac{\abs{u}}{\sqrt{\tilde{g}\,h\,\cos\phi}}\,,
\end{equation*}
where $h$ is the characteristic height of the density current, $u$ is the typical flow velocity and $\phi$ is the characteristic downslope angle. The \textsc{Froude} number is a dimensionless quantity, which expresses the relative importance of buoyancy and inertia.

Finally, the \textsc{Richardson} number can be also introduced \cite{Turner1973}:
\begin{equation*}
  \mathrm{Ri}\ \eqdef\ \frac{1}{\Fr_{\,\mathrm{d}}^{\,2}}\ =\ \frac{\tilde{g}\,h\,\cos\phi}{u^{\,2}}\,.
\end{equation*}
For example, based on extensive experimental observations, \textsc{Keulegan} (1957) \cite{Keulegan1957} derived a simple equation to estimate the head velocity:
\begin{equation*}
  U_{\,f}\ \approx\ C_{\,K}\;\sqrt{\tilde{g}\,h_{\,f}}\,,
\end{equation*}
where $C_{\,K}$ is a constant belonging to the range $\bigl[\,0.7,\,0.9\,\bigr]$ and $h_{\,f}$ is the flow head height.

One could also introduce the so-called \textsc{P\'eclet} number, which measures the relative importance of advective effects to the solute\footnote{It could be also the thermal diffusion in other problems, see \cite{Chhay2015}.} diffusion. However, along with the \textsc{Reynolds} number this parameter is extremely large and, thus, of little help.

Early mathematical models used simple empirical relations to parametrize mixing across the interfaces. These relations have typically been based on the \textsc{Richardson} number \cite{Ellison1959}. Recently, a simple shallow water model was augmented with an empirical experiment-based entrainment rate relation involving the \textsc{Richardson} number \cite{Johnson2013a}. Taking into account that \textsc{Reynolds} and \textsc{P\'eclet} numbers are huge for turbidity currents, the \textsc{Richardson} number is the only primary parameter for this type of flows. We note also that below we use rather its \emph{alter ego} --- the \textsc{Froude} number.

This remark partially explains the importance of various dimensionless numbers in the modelling of turbidity density currents.
\end{remark}

The flow under consideration is clearly turbulent and we are going to take into account this fact into our model. Namely, we apply the classical \textsc{Reynolds} decomposition of all the fields present in our model into the mean and fluctuating parts \cite{Sreenivasan1999, Grinstein2002, Alfonsi2009}:
\begin{equation*}
  \u\ \longrightarrow\ \u\ +\ \u^{\,\prime}, \qquad p\ \longrightarrow\ p\ +\ p^{\,\prime}, \qquad b\ \longrightarrow\ b\ +\ b^{\,\prime}\,,
\end{equation*}
where for the sake of compactness we denote the average part by the same symbol, while the fluctuations are denoted with primes. After substituting this decomposition into the governing Equations \eqref{eq:bouss1} -- \eqref{eq:bouss2} and applying an averaging operator \cite{Lumley2001, Sreenivasan1999}, we obtain the following \textsc{Reynolds} averaged version of the model:
\begin{eqnarray}\label{eq:turb1}
  \div\u\ &=&\ 0\,, \\
  b_{\,t}\ +\ \div[\,b\,\ut\ +\ \overline{b^{\,\prime}\,\u^{\,\prime}}\,]\ &=&\ 0\,, \\
  \u_{\,t}\ +\ \div\bigl[\,\u\otimes\u\ +\ \overline{\u^{\,\prime}\otimes\u^{\,\prime}}\ +\ \frac{p}{\rho_a}\;\Id\,\bigr]\ &=&\ \b\,,
\end{eqnarray}
where the over bar denotes the classical \textsc{Reynolds} average \cite{Sagaut2006}. The same decomposition and averaging operations can be applied to the energy conservation Equation \eqref{eq:energy}:
\begin{multline}\label{eq:kturb}
  \bigl(\half\,(\abs{\u}^2\ +\ q^2)\ +\ b\,y\bigr)_{\,t}\ +\ \div\Bigl[\,\Bigl(\half\,(\,|\u|^2\ +\ q^2)\ +\ b\,y\Bigr)\,\u\ +\ \frac{p}{\rho_0}\;\Id\,\Bigr]\ + \\ 
  +\ \div\Bigl[\,\overline{\u'\otimes\u'\,}\,\u\ +\ y\,\overline{b^{\,\prime}\,\u^{\,\prime}\,}\ 
  +\ \frac{\overline{p^{\,\prime}\,\u^{\,\prime}\,}}{\rho_a}\ +\ \overline{\abs{\u^{\,\prime}}^2\,\u^{\,\prime}}\,\Bigr]\ =\ y\,(b\,v_s)_y\,,
\end{multline}
where $q\ \eqdef\ \abs{\u^{\,\prime}}$ is the mean squared velocity of fluctuations\footnote{The quantity $q$ should not be confused with the specific turbulent kinetic energy, which is proportional to $q^{\,2}$ in our notation \cite{Mohammadi1994}.} \cite{Mohammadi1994, Sagaut2006}. We underline that up to now, besides the \textsc{Boussinesq} assumption \eqref{eq:bouss} no other simplifications have been undertaken.

\begin{remark}
Despite the fact that we consider a two-dimensional problem, the two-dimensional turbulence is unstable in the sense that a tiny perturbation quickly leads to a fully three-dimensional turbulent state. We would like to underline that Equation \eqref{eq:kturb} is valid in this case as well. In fact, we can assume that the velocity field $\u\ =\ (u,\,v,\,\mathrm{w})$ is such that the mean transverse component $w$ is equal identically to zero everywhere, while the averaged fluctuations speed $q$ incorporates three components of the velocity field:
\begin{equation*}
  q\ \eqdef\ \abs{\u^{\,\prime}}\ \eqdef\ \sqrt{(u^{\,\prime})^2\ +\ (v^{\,\prime})^2\ +\ (\mathrm{w}^{\,\prime})^2}\,.
\end{equation*}
Consequently, we consider a two-dimensional flow embedded into a three-dimensional turbulence state.
\end{remark}

In the flow configuration that we consider in the present study, the horizontal momentum is advected vertically by the \textsc{Reynolds} stress $\tau\ =\ -\,\rho_{\,0}\,\overline{u^{\,\prime}\,v^{\,\prime}}\,$. In the developed turbulent flow, the following closure relations have been confirmed experimentally \cite{Townsend1980}:
\begin{equation*}
  \overline{u^{\,\prime}\,v^{\,\prime}\,}\ =\ -\,\tilde{\sigma}\, q^2\,, \qquad \tilde{\sigma}\ \eqdef\ \sigma\,\sign\,(u_{\,y})\,,
\end{equation*}
where $\sigma$ is a positive constant which will be considered below as a small parameter. Typically, the value $\sigma\ \approx\ 0.15$ corresponds fairly well to the experimental observations. Moreover, the following assumptions are generally adopted \cite{Townsend1980}:
\begin{equation*}
  \overline{b^{\,\prime}\,v^{\,\prime}\,}\ =\ -\tilde{\sigma}\,q\,\theta\,, \qquad \mathrm{where} \qquad
  \theta^{\,2}\ \eqdef\ \overline{(b^{\,\prime})^{\,2}}\,.
\end{equation*}
The turbulence is traditionally assumed to be isotropic, \ie
\begin{equation}\label{eq:isotrop}
  \overline{(u^{\,\prime})^2}\ \equiv\ \overline{(v^{\,\prime})^{\,2}}\,.
\end{equation}
Finally, the third correlations in Equation \eqref{eq:kturb} are supposed to have the following asymptotic behaviour:
\begin{equation*}
  \overline{\half\,(u^{\,\prime})^3\ +\ \half\, u^{\,\prime}\,(v^{\,\prime})^{\,2}\ +\ \frac{p^{\,\prime}\,u^{\,\prime}}{\rho_{\,0}}\,}\ =\ 
  o(\sigma)\,q^{\,3}\,,
  \qquad
  \overline{\half\;(u^{\,\prime})^{\,2}\,v^{\,\prime}\ +\ \half\;(v^{\,\prime})^3\ +\ \frac{p^{\,\prime}\,v^{\,\prime}}{\rho_{\,0}}\,}\ =\ 
  o(\sigma)\,q^{\,3}\,.
\end{equation*}

\begin{remark}
We note that the isotropy condition \eqref{eq:isotrop} can be relaxed without any change to the subsequent derivation. Namely, we can assume a weaker condition on the averages of quadratic velocity fluctuations:
\begin{equation*}
  \overline{(u^{\,\prime})^{\,2}}\ \approx\ \overline{(v^{\,\prime})^{\,2}}\,,
\end{equation*}
where we allow for small deviations from the isotropy property which can be of the same order than the small parameter $\sigma\,$:
\begin{equation*}
  \overline{(u^{\,\prime})^{\,2}}\ -\ \overline{(v^{\,\prime})^{\,2}}\ \propto\ \sigma\cdot q^{\,2}\,.
\end{equation*}
\end{remark}

\begin{remark}
The quantity $\theta$ represents the magnitude of mean buoyancy fluctuations. It can be determined from the following equation:
\begin{equation*}
  \bigl(\half\, b^{\,2}\ +\ \half\;\theta^{\,2}\bigr)_{\,t}\ 
  +\ \Bigl[\,\bigl(\half\, b^{\,2}\ +\ \half\,\theta^{\,2}\bigr)\,u\,\Bigr]_{\,x}\ +\ \Bigl[\,\bigl(\half\, b^{\,2}\ +\ \half\,\theta^{\,2}\bigr)\,v\ +\ b\;\overline{b^{\,\prime}\,v^{\,\prime}\,}\,\Bigr]_{\,y}\ =\ 0\,,
\end{equation*}
where subscripts ${}_{x,\,y}$ denote differentiation with respect to those variables. Since the quantity $\theta$ does not explicitly appear in the final model, we do not include this equation into our consideration. However, this equation is needed, for example, if one is interested in reconstructing the vertical structure of the flow \cite[Chapter~7]{Liapidevskii2000}. This issue will be addressed in future studies.
\end{remark}

\subsection{Long wave scaling}

In order to simplify further the governing Equations \eqref{eq:turb1} -- \eqref{eq:kturb} we apply the classical long wave scaling of independent and dependent variables (this operation can be also seen as a passage to thoroughly chosen dimensionless variables):
\begin{equation*}
  \left\{
  \begin{array}{l}
    x\ \rightarrow\ x\,, \qquad y\ \rightarrow\ \eps y\,, \quad t \rightarrow \eps^{-\frac12}\, t\,, \qquad \sigma\ \rightarrow\ \eps\,\sigma\,, \\
    u\ \rightarrow\ \eps^\frac12\, u\,, \qquad v\ \rightarrow\ \eps^\frac32\, v\,, \qquad v_s\ \rightarrow\ \eps^\frac32\, v_s\,, \qquad p\ \rightarrow\ \eps\,\rho_0\, p\,, \qquad b\ \rightarrow\ b\,, \\
    u^{\,\prime}\ \rightarrow\ \eps^\frac12\, u^{\,\prime}\,, \qquad v^{\,\prime}\ \rightarrow\ \eps^\frac12\, v^{\,\prime}\,, \qquad p^{\,\prime}\ \rightarrow\ \eps \rho_0\, p^{\,\prime}\,, \quad b^{\,\prime}\ \rightarrow\ b^{\,\prime}\,,
  \end{array}
  \right.
\end{equation*}
where $\eps\ \eqdef\ \bigl(\frac{h_{\,0}}{\ell}\bigr)^{\,2}\ \ll\ 1$ is the shallowness parameter related to the aspect flow ratio, $h_{\,0}$ being the characteristic flow depth and $\ell$ the typical wavelength. After applying this rescaling to Equations \eqref{eq:turb1} -- \eqref{eq:kturb} and neglecting the higher order terms in $\eps$ we obtain the following system of equations:
\begin{eqnarray}\label{eq:stretch1}
  u_{\,x}\ +\ v_{\,y}\ &=& 0\,, \\
  b_{\,t}\ +\ [\,b\,u\,]_{\,x}\ +\ \bigl[\,b\,(v\ -\ v_{\,s})\ +\ \eps^{\,-1}\;\overline{b^{\,\prime}\,v^{\,\prime}\,}\,\bigr]_{\,y}\ &=& 0\,, \\
  u_{\,t}\ +\ \bigl[\,u^{\,2}\ +\ \P\,\bigr]_{\,x}\ +\ \bigl[\,u\,v\ +\ \eps^{\,-1}\;\overline{u^{\,\prime}\,v^{\,\prime}\,}\,\bigr]_{\,y}\ &=& 0\,, \\
  \bigl(\half\, (u^{\,2}\ +\ q^{\,2})\ +\ b\,y\bigr)_{\,t}\ +\ \bigl[\,\half\,(u^{\,2}\ +\ q^{\,2})\,u\ +\ b\,y\,u\ +\ \P\,u\,\bigr]_{\,x}\ +\ && \nonumber \\ 
  \bigl[\,\half\,(u^{\,2} + q^{\,2})\,v\ +\ b\,y\,(v\ -\ v_{\,s})\ +\ \P\,v\,\bigr]_{\,y}\ +\ && \nonumber \\
  \eps^{\,-1}\;[\,u\,\overline{u^{\,\prime}\,v^{\,\prime}\,} + y\,\overline{b^{\,\prime}\,v^{\,\prime}\,}\,]_{\,y}\ &=& -\,b\,v_{\,s}\,.
\end{eqnarray}
The pressure $\P$ is defined by the following relations:
\begin{equation*}%\label{eq:stretch2}
  \P_{\,y}\ =\ -b\,, \qquad \P\ \eqdef\ p\ +\ \overline{(u^{\,\prime})^{\,2}}\ \equiv\ p\ +\ \overline{(v^{\,\prime})^{\,2}}\,,
\end{equation*}
where we used the fluctuations isotropy property \eqref{eq:isotrop}. The first relation shows that the pressure $\P$ is hydrostatic. The second relation says that it is not the modified pressure $p\,$, which is hydrostatic, but its combination with a component $\overline{(u^{\,\prime})^{\,2}}$ (or equivalently $\overline{(v^{\,\prime})^{\,2}}$) of the \textsc{Reynolds} stress. Hereafter we will return to dimensional variables since the long wave scaling was already applied to the governing equations. We mention also that an alternative derivation for a similar depth-averaged system can be found in \cite[Appendix~A]{Johnson2013a}.

\begin{figure}
  \centering
  \includegraphics[width=0.89\textwidth]{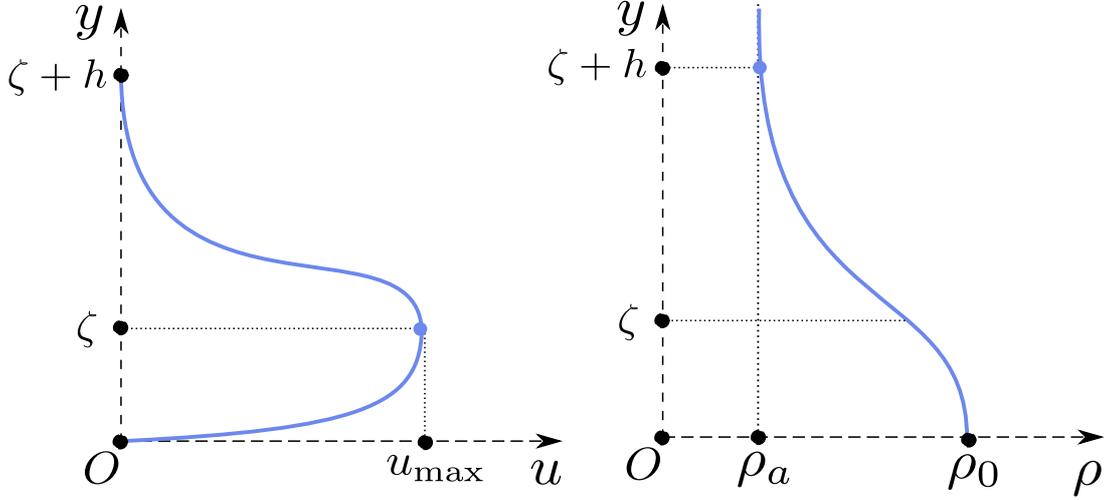}
  \caption{\small\em Vertical flow structure in turbidity currents. On the left image we schematically represent the qualitative behaviour of the horizontal velocity variable $u\,$, while on the right image we show the behaviour of the density $\rho$. The vertical coordinate $y$ is measured from the solid bottom $d(x)$. This schematic representation is supported by findings reported in \cite{Pawlak1998}.}
  \label{fig:struct}
\end{figure}

The vertical structure of turbidity gravity flows has been investigated experimentally \cite{Alavian1986, Garcia1993a, Pawlak1998}. The horizontal velocity and density dependence on the flow depth are schematically represented in Figure~\ref{fig:struct}. The density $\rho$ is obviously constant in the bottom ($\rho_{\,0}$) and pure water ($\rho_{\,a}$) layers. In the mixing layer the density varies continuously and almost linearly between two constant boundary values. The horizontal velocity $u\,(\x,\,t)$ attains its maximum on the boundary between the bottom and mixing layers. Then the velocity $u\,(\x,\,t)$ goes to zero value on the solid bottom and on the boundary with the still water. The horizontal velocity $u\,(\x,\,t)$ behaves in the bottom as in a turbulent boundary layer \cite{Bradshaw1967, Townsend1980, Liu2006a}. These experimental evidences suggest us to apply a multi-layer approximation to simplify the model while resolving the vertical structure of the flow.

Consequently, using this a priori knowledge about the flow structure, we will apply the classical vertical averaging operator to obtain a depth-integrated model. We do not provide here the details about these computations, however we recall some basic assumptions we need to adopt:
\begin{itemize}
  \item The inertial turbulence scale $\ell_{\,T}$ in the mixing layer is determined by large eddies and thus, it is comparable to the layer depth: $\ell_{\,T}\ \propto\ h\,$.
  \item In the upper quiescent layer, by definition, the water is still (without sediment particles) and it is motionless.
  \item On the solid boundary $y\ =\ d\,(x)$ we have the usual bottom impermeability boundary condition. The \textsc{Reynolds} stresses at the bottom are given by the following relations:
  \begin{equation*}
    \left.\abs{\overline{u^{\,\prime}v^{\,\prime}\,}}\right|_{\,y\,=\, d\,(x)}\ \equiv\ u_{\,\ast}^{\,2}\ =\ c_{\,w}\,w\,\abs{w}\ \equiv\ c_{\,w}\,w^{\,2}\,, \qquad w\ \geq\ 0\,,
  \end{equation*}
  \begin{equation*}
    \left.\overline{b^{\,\prime}v^{\,\prime}\,}\right|_{y\,=\,d\,(x)}\ =\ 0\,.
  \end{equation*}
  The variable $u_{\,\ast}$ is the so-called friction velocity and the quantity $u_{\,\ast}^{\,2}$ is proportional to the turbulent friction stress\footnote{The drag (or friction) associated with bottom must be parametrized since shallow water depth-averaged equations do not resolve turbulence produced at this boundary. This is commonly achieved by introducing a drag coefficient along with an empirical law for the shear velocity near the bottom.} at the solid boundary and $c_{\,w}\ \in\ \R^{+}$ is a (positive) constant.
  \item The sedimentation speed $v_{\,s}\ =\ 0$ is equal to zero and the buoyancy $b\,(\x,\,t)\ \equiv\ b_{\,0}$ is constant in the bottom layer $d\,(x)\ <\ y\ <\ d\,(x)\ +\ \zeta\,(x,\,t)\,$.
\end{itemize}
The buoyancy was defined in \eqref{eq:buo}, whose value $b_{\,0}$ in the bottom layer is equal to
\begin{equation}\label{eq:b0}
  b_{\,0}\ \equiv\ \frac{\bigl(\rho_{\,0}\ -\ \rho_{\,a}\bigr)\,g}{\rho_{\,a}}\,.
\end{equation}
The depth-averaged velocities $u\,$, $w$ and densities $\rho_0\,$, $\rhob$ are defined in Figure~\ref{fig:sketch} and summarized in Table~\ref{tab:defs}. We reiterate on the fact that the sediments porosity can be taken into account in the present modelling by choosing the sediment concentration $\pi\,(\x,\,t)$ (appearing in $b_{\,0}$) smaller than one. More complicated approaches exist as well \cite{Meyer-Peter1948, Grass1981, Nielsen1992, MoralesdeLuna2009, Benkhaldoun2009, Benkhaldoun2009a}.

\begin{table}
  \centering
  \begin{tabular}{c|l}
  \hline\hline
  \textit{Variable} & \textit{Significance} \\
  \hline\hline
  $b_{\,s}$ & Buoyancy of sediments \\
  $b_{\,0}$ & Buoyancy of the lower (bottom) layer \\
  $b\,(x,\,t)$ & Buoyancy in the mixing layer \\
  $h_{\,s}$ & Thickness of the sediment deposit \\
  $\zeta\,(x,\,t)$ & Thickness of the lower (bottom) layer \\
  $h\,(x,\,t)$ & Thickness of the mixing layer \\
  $w\,(x,\,t)$ & Depth-averaged mean velocity in the bottom layer \\
  $u\,(x,\,t)$ & Depth-averaged mean velocity in the mixing layer \\
  $q\,(x,\,t)$ & Depth-averaged turbulent mean squared velocity in the mixing layer \\
  $\eta\,(x,\,t)$ & Thickness of the ambient fluid layer \\
  $v\,(x,\,t)$ & Depth-averaged velocity in the ambient fluid \\
  \hline\hline
  \end{tabular}
  \bigskip
  \caption{\small\em Definitions of various parameters used in the approximate models. For the illustration see Figure~\ref{fig:sketch}. A more complete list of employed nomenclature is given in Appendix~\ref{app:nom}.}
  \label{tab:defs}
\end{table}

After performing the depth-averaging operation under the incompressibility condition \eqref{eq:stretch1} and all the assumptions listed above, we obtain the mass, momentum and energy conservation laws for the averaged quantities:
\begin{equation}\label{eq:massC}
  (b_0\,\zeta\ +\ b\,h)_{\,t}\ +\ [\,b_{\,0}\,\zeta\,w\ +\ b\,h\,u\,]_{\,x}\ =\ 0\,,
\end{equation}
\begin{multline}\label{eq:momentC}
  (\zeta\,w\ +\ h\,u)_{\,t}\ +\ \bigl[\,\zeta\,w^{\,2}\ +\ h\,u^{\,2}\ +\ \half\, b_{\,0}\,\zeta^{\,2}\ +\ b\,\zeta\,h\ +\ \half\, b\,h^{\,2}\,\bigr]_{\,x}\\ =\ -\,(b_{\,0}\,\zeta\ +\ b\,h)\,d_{\,x}\ -\ u_{\,\ast}^{\,2}\,,
\end{multline}
\begin{multline}\label{eq:energyC}
  \bigl(\zeta\,w^{\,2}\ +\ h\,(u^{\,2}\ +\ q^{\,2})\ +\ b_{\,0}\,\zeta^{\,2}\ +\ 2\,b\,\zeta\,h\ +\ b\,h^{\,2}\bigr)_{\,t}\ + \\
  \bigl[\,\zeta\,w^{\,3}\ +\ h\,u\,(u^{\,2}\ +\ q^{\,2})\ +\ 2\,b_{\,0}\,\zeta^{\,2}\,w\ +\ 2\,b\,\zeta\,h\,w\ +\ 2\,b\,(\zeta\ +\ h)\,h\,u\,\bigr]_{\,x}\ =\\ -\E\ -\ 2\,(b_{\,0}\,\zeta\,w\ +\ b\,h\,u)\,d_{\,x}\ -\ 2\,w\,u_{\,\ast}^{\,2}\,,
\end{multline}
where we introduce into the model a dissipative term $\E$ coming from the turbulent energy dissipation \cite{Sreenivasan1999, Lesieur2008}. The form of this dissipative term has to be specified. The bottom stress (or friction) term $\tau_{\,\ast}$ was already specified above. We remind that in the high \textsc{Reynolds} number regime, the following expression is widely used:
\begin{equation*}
  u_{\,\ast}^{\,2}\ =\ c_{\,w}\,w^{\,2}\,.
\end{equation*}
For the sediments layer we can write a separate momentum balance equation after the same averaging procedure:
\begin{equation}\label{eq:uplus}
  w_{\,t}\ +\ \bigl[\,\half\,w^2\ +\ b_0\,\zeta\ +\ b\,h\,\bigr]_{\,x}\ =\ -b_0\,d_{\,x}\ -\ \frac{u_{\,\ast}^{\,2}}{\zeta}\,,
\end{equation}
provided that the layer height $\zeta\ >\ 0\,$. These equations have to be completed by two kinematic conditions on the interfaces:
\begin{align}\label{eq:kin1}
  \zeta_{\,t}\ +\ [\,\zeta\,w\,]_{\,x}\ &=\ \chi^{\,-}\,, \\
  h_{\,t}\ +\ [\,h\,u\,]_{\,x}\ &=\ \chi^{\,+}\,,\label{eq:kin2}
\end{align}
where the terms $\chi^{\,\pm}$ are the entrainment rates which account for the mass exchanges between the layers. The mass exchange between the mixing and ambient layers is driven by the widely known \textsc{Kelvin}--\textsc{Helmholtz} instability \cite{Helmholtz1868}. This suggests also the presence of an underlying coupled hydrodynamic/sediment layer instability as well.

Using Equations \eqref{eq:uplus}, \eqref{eq:kin1} and \eqref{eq:kin2} the just derived conservation laws \eqref{eq:massC} -- \eqref{eq:energyC} can be rewritten also in the following non-conservative form (the whole system is listed here):
\begin{align}\label{eq:nc1}
  \zeta_{\,t}\ +\ [\,\zeta\,w\,]_{\,x}\ =&\ \chi^{\,-}\,, \\
  h_{\,t}\ +\ [\,h\,u\,]_{\,x}\ =&\ \chi^{\,+}\,, \label{eq:nc2} \\
  w_{\,t}\ +\ w\,w_{\,x}\ +\ [\,b_{\,0}\,\zeta\ +\ b\,h\,]_{\,x}\ =&\ -b_{\,0}\,d_{\,x}\ -\ \frac{u_{\,\ast}^{\,2}}{\zeta}\,, \label{eq:nc3} \\
  b_{\,t}\ +\ u\,b_{\,x}\ =&\ -\frac{b_{\,0}\,\chi^{\,-}\ +\ b\,\chi^{\,+}}{h}\,, \label{eq:nc4} \\
  u_{\,t}\ + u\,u_{\,x}\ +\ b\,[\,\zeta\ +\ h\,]_{\,x}\ +\ \half\,h\,b_{\,x}\ =&\
  -\ \frac{w\,\chi^{\,-}\ +\ u\,\chi^{\,+}}{h} - b\,d_{\,x}\,, \label{eq:nc5} \\
  q_{\,t}\ +\ u\,q_{\,x}\ =&\ (2\,h\,q)^{\,-1}\Bigl[\,\bigl(2\,w\,u\ -\ w^{\,2}\ +\ b_{\,0}\,h\ -\ 2\,b\,h\bigr)\,\chi^{\,-}\ + \nonumber \\
  & \bigl(u^{\,2}\ -\ q^{\,2}\ -\ b\,h\bigr)\,\chi^{\,+}\ -\ \E\ -\ b\,h\,v_{\,s}\,\Bigr]\,.\label{eq:nc6}
\end{align}
In particular, this computation shows that our system possesses multiple contact characteristics $\odd{x}{t}\ =\ \upxi_{\,3,\,4}\ =\ u$ (the indices $3$ and $4$ will become clearer below in Section~\ref{sec:sed}). Below we will use this property to extract from this model a subsystem which governs the mixing layer dynamics. At the current stage, disregarding the particular form of the mixing terms, our model has the structure of a two-layer shallow water system coupled with two advection equations: one for the buoyancy and another one describes the transport of the turbulent averaged velocity fluctuations. In order to close the system above, we have to specify the entrainment rates $\chi^{\,\pm}$ among fluid layers and the energy dissipation term $\E\,$. Some approaches to determine entrainment rates are explained below. For the dissipative term, we assume the following closure relation:
\begin{equation*}
  \E\ \eqdef\ \kappa\,q^{\,3}\,, \qquad \mbox{ where } \qquad \kappa\ >\ 0\,.
\end{equation*}

%%% ----------------------------------------------------------------------- %%%

\subsection{Determination of entrainment rates without sedimentation}
\label{sec:close1}

If the sedimentation velocity $v_{\,s}\ \equiv\ 0\,$, the main hypothesis that we can make about the entrainment rates $\chi^{\,\pm}$ is that they are related to the mean square root of the turbulent velocity $q\,$. The simplest dependence is the linear proportionality and we adopt it in our study:
\begin{equation}\label{eq:close1}
  \chi^{\,\pm}\ \propto\ \sigma^{\,\pm}\,q\,.
\end{equation}
Depending on the flow type, which is realized in practice, the closure law \eqref{eq:close1} can be further refined. We may distinguish between two following situations:
\begin{itemize}
  \item The mass transfer takes place at upper and lower boundaries of the turbulent layer (the so-called ``mixing layer'' described in \cite[Section~3.1]{Liapidevskii2018})
  \item The mass transfer takes place only at the upper boundary (the so-called bottom turbulent jet)
\end{itemize}
So, a more accurate closure relation for mixing layers is obtained by taking $\sigma^{\,+}\ \equiv\ 2\,\sigma\,$, $\sigma^{\,-}\ \equiv\ -\,\sigma\,$, yielding the following expressions for entrainment rates:
\begin{equation}\label{eq:ml}
  \chi^{\,+}\ =\ 2\,\sigma\,q\,, \qquad
  \chi^{\,-}\ =\ -\,\sigma\,q\,.
\end{equation}
For bottom turbulent jets we propose the following closure $\sigma^{\,+}\ \equiv\ \sigma\,$, $\sigma^{\,-}\ \equiv\ 0\,$, which yields
\begin{equation}\label{eq:bott}
  \chi^{\,+}\ =\ \sigma\,q\,, \qquad
  \chi^{\,-}\ =\ 0\,.
\end{equation}
The constant $\sigma$ is usually taken to be in the segment $\bigl[\,0.15,\, 0.17\,\bigr]$ and they characterizes the ratio of the characteristic vertical to horizontal flow scales in the long wave approximation. In the case of flat bottom without friction, the constant $\sigma$ can be completely removed from equations by a suitable scaling of independent variables. We introduce also a new constant $\delta$ as the ratio of two previously introduced constants:
\begin{equation}\label{eq:delta}
  \delta\ \eqdef\ \frac{\kappa}{\sigma}\ >\ 0\,.
\end{equation}
This constant $\delta$ shall be used in developments below. In practice, $\delta$ is taken in the range
\begin{equation*}
  0\ \leq\ \delta\ \apprle\ 8.0\,.
\end{equation*}
We observed using numerical means that $\delta$ has a weak influence on the qualitative properties of unsteady simulations. We would like to mention also that a number of authors considered simple constant entrainment rate coefficients independent of \textsc{Richardson} number \cite{Bonnecaze1999, Turner1986, Tickle1996, Ross2006}.

%%% ----------------------------------------------------------------------- %%%

\subsection{Determination of entrainment rates with sedimentation}

\begin{figure}
  \centering
  \includegraphics[width=0.45\textwidth]{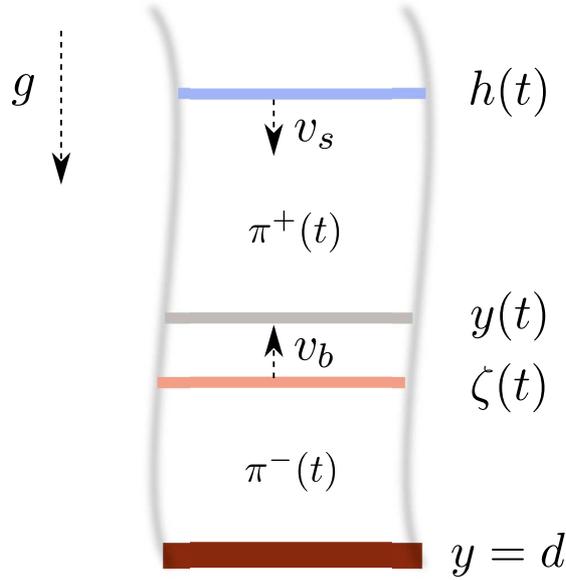}
  \caption{\small\em Schematic illustration to the sedimentation process of a uniform layer in the horizontal extent.}
  \label{fig:entr}
\end{figure}

In order to prescribe the expressions for $\chi^{\pm}$ in terms of other physical quantities, we analyze an idealized situation schematically depicted in Figure~\ref{fig:entr}. Namely, we consider a flat bottom, the horizontal motion is absent and the flow is invariant under the horizontal translations. In other words, we study the evolution of this layered system under the sedimentation dynamics with a constant fall speed $v_s\,$. We denote by $y\,(t)$ the supposed upper boundary of particles which will become a part of the sediment layer at the next time moment $t\ +\ \Delta t\,$. The upper boundary $h\,(t)$ will fall with the speed $v_{\,s}\,$, while the lower layer will grow with some velocity $v_{\,b}$ which is yet to be determined. By definition, the entrainment rates $\chi^\pm$ can be expressed in terms of $v_{\,s}$ and $v_{\,b}$ as follows:
\begin{equation*}
  \chi^+\ \eqdef\ v_{\,s}\ -\ v_{\,b}\,, \qquad \chi^-\ \eqdef\ v_{\,b}\,.
\end{equation*}
In order to determine the speed $v_{\,b}$ we will write two kinematic conditions and one sediments mass conservation equation:
\begin{align*}
  \zeta\,(t\ +\ \Delta t)\ -\ \zeta\,(t)\ =&\ v_{\,b}\cdot\Delta t\,, \\
  y\,(t)\ -\ \zeta\,(t\ +\ \Delta t)\ =&\ v_{\,s}\cdot\Delta t\,, \\
  \bigl(\zeta\,(t\ +\ \Delta t)\ -\ \zeta\,(t)\bigr)\,\pi^{\,-}\,\rho_{\,0}\ =&\ \bigl(y\,(t)\ -\ \zeta\,(t)\bigr)\,\pi^{\,+}\,\rho\,.
\end{align*}
By solving these relations, one can easily find the following expression for the sediments layer growth velocity $v_{\,b}\,$:
\begin{equation*}
  v_{\,b}\ \equiv\ \frac{v_{\,s}\cdot\pi^{\,+}}{\pi^{\,-}\ -\ \pi^{\,+}}\,.
\end{equation*}
Finally, in the presence of the turbulent flow we complete the just derived expression of $\chi^{\,+}$ by a term $\sigma\,q$ responsible of the still water entrainment into the mixing layer. Thus, the final expressions are
\begin{equation}\label{eq:rates}
  \chi^{\,-}\ =\ \frac{v_{\,s}\cdot b}{b_0\ -\ b}\ \equiv\ v_{\,b}\,, \qquad
  \chi^{\,+}\ =\ -\frac{v_{\,s}\cdot b_{\,0}}{b_{\,0}\ -\ b}\ +\ \sigma\,q\,,
\end{equation}
where we replaced equivalently the volume fractions $\pi^{\,\pm}$ by buoyancy variables $b_{\,0}\,$, $b\,$, since the system is written using these variables. As one can see, in the turbulent jet we have two competing effects: the sedimentation of particles down to the bottom and the pure water entrainment into the turbulent layer.

%%% ----------------------------------------------------------------------- %%%

\subsubsection{Simplified model}

Entrainment rate closure \eqref{eq:rates} can be further simplified if we notice that the density of entrained sediments is much lower than sediments layer in the vicinity of the solid bottom. In other words, we can assume that $b\ \ll\ b_0\,$. Under this assumption Equation \eqref{eq:rates} becomes
\begin{equation}\label{eq:vs}
  \chi^{\,-}\ \equiv\ 0\,, \qquad \chi^{\,+}\ =\ \sigma\,q\ -\ v_{\,s}\,.
\end{equation}
The main difference with the closure proposed in Section~\ref{sec:close1} is that $\chi^{\,-}$ vanishes and $\chi^{\,+}$ takes into account the sedimentation velocity $v_{\,s}\,$. Notice also that the sedimentation velocity parameter $v_{\,s}$ may assume eventually zero value and in this way we recover the previously proposed closure relation~\eqref{eq:bott}. This situation is not excluded. This closure turns out to be useful if we aim mostly to describe the dynamics of the mixing layer and of bottom buoyancy jets. In particular, it is successfully used to explain some experiments from \cite{Rastello2004}.

%%% ----------------------------------------------------------------------- %%%

\subsubsection{Intermediate conclusions}

Let us summarize the developments made so far. The proposed model is self-consistent and depends essentially on the values of three constants $\sigma\,$, $\delta$ and $c_{\,w}\,$, which are to be specified before the simulations can be run. Our model covers also the situation where the intermediate mixing layer reaches the bottom and transforms into the so-called bottom buoyancy jet. It happens when $\zeta\ \ll\ h\,$. In this case one has to set $\chi^{\,-}\ =\ 0$ and the computations can be continued. We are aware of the fact that the mixing layer cannot ``touch'' the bottom \emph{stricto sensu}. There is always a little layer of a heavy fluid with density close to $\rho_{\,0}\,$. However, this layer\footnote{It is depicted in Figure~\ref{fig:sketch} under the dense current and it has the velocity $w\,(x,\,t)$ along with the thickness $\zeta\,(x,\,t)\,$.} is dynamically passive since the gravity and bottom friction forces are in equilibrium. In situations, where the mixing layer does not approach bottom (yet), this tiny layer plays the r\^ole in alimenting the flow head with new sediment particles. Moreover, the flow head can be supplied by sediments of density $\rho_{\,0}$ located immediately after the front in the downstream direction since the mixing process is most significant close to the front of the gravity current. Some mathematical considerations on this situation are given in the following Section.

%%% ----------------------------------------------------------------------- %%%

\subsection{Sediments equilibrium model}
\label{sec:sed}

Hereafter we consider a special regime where the flow attains the equilibrium state in the lowest bottom layer. Namely, we assume that the gravity force balances exactly the turbulent friction at the solid bottom, \ie
\begin{equation}\label{eq:frict}
  \frac{c_{\,w}\,w^{\,2}}{\zeta}\ \equiv\ -b_{\,0}\,d_{\,x}\,.
\end{equation}
In other words, the right-hand side of Equation \eqref{eq:uplus} vanishes and this relation provides us an expression of the velocity $w$ as a function of the bottom layer thickness $\zeta$ and of the local bottom slope $d_x\,$:
\begin{equation*}
  w\ \equiv\ \psi(\zeta,\, x)\ \eqdef\ \sqrt{-\frac{b_{\,0}\,\zeta\,d_{\,x}}{c_{\,w}}}\,, \qquad d_{\,x}\ \leq\ 0\,,
\end{equation*}
where we choose the positive branch of solutions $w\ >\ 0\,$, since the flow is expected to move downwards. The bottom layer thickness can be determined by solving the corresponding kinematic condition \eqref{eq:kin1}:
\begin{equation}\label{eq:hplus}
  \zeta_{\,t}\ +\ \bigl[\,\zeta\cdot\psi(\zeta,\,x)\,\bigr]_{\,x}\ =\ \chi^{\,-}\,.
\end{equation}
The last equation is related to the rest of the system only through the entrainment rate $\chi^{\,-}$ at the right-hand side. Thus, the rest of the \emph{equilibrium} system reads:
\begin{align}\label{eq:psi1}
  h_{\,t}\ +\ [\,h\,u\,]_x\ &=\ \chi^{\,+}\,, \\
  u_{\,t}\ +\ u\,u_x\ +\ b\,h_{\,x}\ +\ \half\,h\,b_{\,x}\ &=\ -b\,d_{\,x}\ -\ b_{\,0}\,\zeta_{\,x}\ -\ \frac{\psi\,\chi^{\,-}\ +\ u\,\chi^{\,+}}{h}\,, \\
  b_{\,t}\ +\ u\,b_{\,x}\ &=\ -\,\frac{b_{\,0}\,\chi^{\,-}\ +\ b\,\chi^{\,+}}{h}\,, \\
  q_{\,t}\ +\ u\,q_{\,x}\ &=\ (2\,q\,h)^{-1}\;\bigl[\,(2\,\psi\,u\ -\ \psi^{\,2}\ +\ b_{\,0}\,h\ -\ 2\,b\,h)\,\chi^- \\
  &\ +\ (u^{\,2}\ -\ q^{\,2}\ -\ b\,h)\,\chi^+\ -\ \kappa\,q^{\,3}\ -\ b\,h\,v_{\,s}\,\bigr]\,.\label{eq:psi2}
\end{align}
The last system is always hyperbolic\footnote{In order to check it one has to write it in conservative variables, compute the \textsc{Jacobian} matrix of the advective flux with respect to conservative variables and find its eigenvalues \cite{Lax1973}. We skip these standard steps for the sake of brevity of this study.} with characteristic speeds $\bigl\{\,\upxi_{\,i}\,\bigr\}_{\,i\,=\,1}^{\,5}$ which can be easily computed:
\begin{align*}
  \od{x}{t}\ &=\ \upxi_{\,1,\,2}\ =\ u\ \pm\ \sqrt{b\,h}\,, \\
  \od{x}{t}\ &=\ \upxi_{\,3,\,4}\ =\ u\,, \\
  \od{x}{t}\ &=\ \upxi_{\,5}\ =\ \psi\ +\ \pd{\psi}{\zeta}\;\zeta\,.
\end{align*}
The last characteristic $\upxi_{\,5}$ corresponds to kinematic waves in the lowest bottom layer described by Equation \eqref{eq:hplus}.

%%% ----------------------------------------------------------------------- %%%

\subsection{Equilibrium model without sedimentation}

The system derived in the previous Section can be further simplified if we neglect the effect of the sedimentation velocity $v_{\,s}\,$, which is possible if $\abs{v_{\,s}}\ \ll\ q\,$. In simple words, hereafter we will set this parameter to zero:
\begin{equation}\label{eq:vs0}
  v_{\,s}\ \equiv\ 0\,.
\end{equation}
This approximation is valid in a fully developed turbulent flow on time scales comparable to the lifetime of large structures (the so-called inertial range). It can be also seen from a different physical perspective: the sedimentation can be neglected while the transport of sediment particles is governed by the main eddies of the gravity current over an incline. Of course, it has to be taken into account in regions where the suspension particles trajectory looks like a free fall. Under the assumption \eqref{eq:vs0}, the entrainment rates $\chi^\pm$ according to \eqref{eq:vs} become
\begin{equation*}
  \chi^{\,-}\ \equiv\ 0\,, \qquad \chi^{\,+}\ \equiv\ \sigma\,q\,,
\end{equation*}
which implies that the kinematic Equation \eqref{eq:hplus} is completely decoupled from System \eqref{eq:psi1} -- \eqref{eq:psi2}. The information about $\zeta$ is transported along characteristics of this equation with the speed $\upxi_{\,5}\,$. So, if $\zeta$ is constant initially and this constant value is maintained at the channel inflow, $\zeta$ will remain so under the system dynamics. We will adopt this assumption as well in order to focus our attention on the mixing layer dynamics. Below we will consider only the middle mixing layer, which was considered in \cite{Liapidevskii2018}. Under these conditions, the equilibrium model \eqref{eq:psi1} -- \eqref{eq:psi2} becomes:
\begin{align}\label{eq:eq1}
  h_{\,t}\ +\ [\,h\,u\,]_{\,x}\ =&\ \sigma\,q\,, \\
  b_{\,t}\ +\ u\, b_{\,x}\ =&\ -\frac{\sigma\, q\, b}{h}\,, \\
  u_{\,t}\ +\ u\,u_{\,x}\ +\ b\,h_{\,x}\ +\ \half\, h\, b_{\,x}\ =&\ -\frac{\sigma\, q\, u}{h}\ -\ b\, d_{\,x}\,, \\
  q_{\,t}\ +\ u\,q_{\,x}\ =&\ \sigma\;\frac{u^{\,2}\ -\ q^{\,2}\ - h\, b\ -\ \delta q^{\,2}}{2\,h}\,,\label{eq:eq4}
\end{align}
where the constant $\delta$ was defined above in \eqref{eq:delta}. The system \eqref{eq:eq1} -- \eqref{eq:eq4} can be equivalently recast in the conservative form which has an advantage to be valid for discontinuous solutions as well \cite{Lax1973, Godlewski1990}:
\begin{align}\label{eq:cons1}
  h_{\,t}\ +\ [\,h\,u\,]_{\,x}\ =&\ \sigma\,q\,, \\
  (h\,b)_{\,t}\ +\ [\,h\,b\,u\,]_{\,x}\ =&\ 0\,, \\
  (h\,u)_{\,t}\ +\ \bigl[\,h\,u^{\,2}\ +\ \half\, b\, h^{\,2}\,\bigr]_{\,x}\ =&\ -h\,b\,d_{\,x}\,, \\
  \bigl(h\,(u^{\,2}\ +\ q^{\,2}\ + h\,b)\bigr)_{\,t}\ +\ \bigl[\,(u^{\,2}\ +\ q^{\,2}\ +\ 2\,h\,b)\;h\,u\,\bigr]_{\,x}\ =&\ -2\,h\,b\,u\,d_{\,x}\ -\ \kappa\,q^{\,3}\,.\label{eq:cons4}
\end{align}
Here, $\kappa\ \in\ \R^{\,+}$ is a positive constant measuring the rate of turbulent dissipation \cite{Sreenivasan1999, Lesieur2008}.

An important property of this model is the absence of the bottom friction term. This physical effect was taken into account in \eqref{eq:frict} while excluding the bottom layer (of thickness $\zeta(x,\,t)$). This gives us the mathematical reason for the absence of a friction term in model \eqref{eq:cons1} -- \eqref{eq:cons4}. Similarly, we can bring also a physical argument to support this fact. The horizontal velocity takes the maximum value on the boundary between the bottom sediment and mixing layers (see Figure~\ref{fig:struct}). Consequently, the \textsc{Reynolds} stress $\tau$ vanishes here.

\begin{remark}
In fact, we can derive an additional balance law by combining together Equations \eqref{eq:eq1} and \eqref{eq:eq4}:
\begin{equation*}
  (h\,q)_{\,t}\ +\ [\,h\,q\,u\,]_{\,x}\ =\ \half\,\sigma\;\bigl(u^{\,2}\ +\ (1\ -\ \delta)\,q^{\,2}\ -\ h\,b\bigr)\,.
\end{equation*}
The last equation is not independent from Equations \eqref{eq:cons1} -- \eqref{eq:cons4}. Consequently, it does not bring new information about the equilibrium sedimentation-free model \eqref{eq:eq1} -- \eqref{eq:eq4}. Nevertheless, we provide it here for the sake of the exposition completeness.
\end{remark}

The proposed model has an advantage of being simple, almost physically self-consistent\footnote{Only two constants $\sigma$ and $\delta$ need to be prescribed to close the system.} and having the hyperbolic structure. It was derived for the first time in \cite[Chapter~5]{Liapidevskii2000}. In order to obtain a well-posed problem the system \eqref{eq:cons1} -- \eqref{eq:cons4} has to be completed by corresponding boundary and initial conditions. Above, in the main parts of this article, the just derived equilibrium System~\eqref{eq:cons1} -- \eqref{eq:cons4} is studied in more details by analytical and numerical means.

%%% ----------------------------------------------------------------------- %%%

\section{Ambient fluid motion}
\label{sec:amb}

Above in our derivation we assumed that the ambient fluid layer is motionless. In order to model various types (overflow, thermals or avalanches) of density currents over a slope taking into account the flow of the ambient fluid, one has to use a three-layer model such as the one proposed in \cite{Liapidevskii2004}. These equations describe more accurately the formation and dynamics of the mixing layer between two homogeneous layers of different densities. This is essentially the result of the hydrodynamic shear instability. In this way, the entrainment processes of the ambient fluid have to be taken into account. Above we presented, for simplicity, the derivation of a two-layer model where the ambient fluid layer was unbounded (from above) and motionless. However, the same derivation procedure can be easily generalized to the channels of finite total depth, which are investigated in the current work. In this Section we neglect the sedimentation velocity (\ie~$v_{\,s}\ \equiv\ 0$) for the sake of simplicity. Under the \textsc{Boussinesq} hypothesis $\dfrac{\rho_{\,a}}{\rho_{\,0}}\ \ll\ 1\,$, the system describing the motion of lower and upper layers takes the form:
\begin{align}\label{eq:sys1}
  \zeta_{\,t}\ +\ \bigl[\,\zeta\,w\,\bigr]_{\,x}\ &=\ -\chi^{\,-}\,, \\
  \eta_{\,t}\ +\ \bigl[\,\eta\,v\,\bigr]_{\,x}\ &=\ -\chi^{\,+}\,, \label{eq:sys2}\\
  w_{\,t}\ +\ \Bigl[\,\half\,w^{\,2}\ +\ b_{\,0}\,\zeta\ +\ b\,h\ +\ p\,\Bigr]_{\,x}\ &=\ -b_{\,0}\,d_{\,x}\ -\ \frac{u^{\,2}_{\,\ast}}{\zeta}\,,\label{eq:sys3} \\
  v_{\,t}\ +\ \bigl[\,\half\,v^{\,2}\ +\ p\,\bigr]_{\,x}\ &=\ 0\,.\label{eq:sys4}
\end{align}
Here $\zeta\,$, $\eta$ and $h$ are thicknesses of the lower, upper and mixing layers correspondingly; $w\,$, $v$ and $u$ are corresponding depth-averaged horizontal velocities. The constant and variable buoyancies are defined as above:
\begin{equation*}
  b_{\,0}\ \eqdef\ \frac{g\,(\rho_{\,0}\ -\ \rho_{\,a})}{\rho_{\,a}}\,, \qquad b\ \eqdef\ \frac{g\,(\rho\ -\ \rho_{\,a})}{\rho_{\,a}}\,.
\end{equation*}
The densities $\rho_{\,0}\,$, $\rho$ and $\rho_{\,a}$ are obviously measured in the lower, mixing and upper layers respectively. The new ingredient in this model is the pressure $p$ applied to the upper boundary of the flow. The entrainment velocities $\chi^{\,\pm}$ are supposed to be known functions of other variables to close the system. Under \textsc{Boussinesq} approximation, the upper boundary is either horizontal:
\begin{equation*}
  \H\ =\ h\ +\ \eta\ +\ \zeta\ =\ \H_{\,0}\ -\ d\,(x)\ \defeq\ \H\,(x)\,, \qquad
  \H_{\,0}\ =\ \const,
\end{equation*}
or is replaced by an inclined rigid lid:
\begin{equation*}
  \H\ =\ h\ +\ \eta\ +\ \zeta\ \equiv\ \H_{\,0}\ =\ \const.
\end{equation*}
The total fluid flux $\Q$ in the channel is a given function of time:
\begin{equation*}
  \Q\,(t)\ \eqdef\ \zeta\,w\ +\ h\,u\ +\ \eta\,v\,.
\end{equation*}
In order to obtain a closed system, we have to complete Equations~\eqref{eq:sys1} -- \eqref{eq:sys3} by the mass, momentum and energy conservation equations:
\begin{equation}\label{eq:sys5}
  \bigl(b_{\,0}\,\zeta\ +\ b\,h\bigr)_{\,t}\ +\ \bigl[\,b_{\,0}\,\zeta\,w\ +\ b\,h\,u\,\bigr]_{\,x}\ =\ 0\,,
\end{equation}
\begin{multline}\label{eq:sys6}
  \Q_{\,t}\ +\ \Bigl[\,\zeta\,w^{\,2}\ +\ h\,u^{\,2}\ +\ \eta\,v^{\,2}\ +\ \half\,b_{\,0}\,\zeta^{\,2}\ +\ b\,h\,\zeta\ +\ \half\,b\,h^{\,2}\ +\ \H\,p\,\Bigr]_{\,x} \\
 =\ -(p\ +\ b_{\,0}\,\zeta\ +\ b\,h)\,d_{\,x}\ -\ u_{\,\ast}^{\,2}\,,
\end{multline}
\begin{multline}\label{eq:sys7}
  \Bigl(\zeta\,w^{\,2}\ +\ h\,(u^{\,2}\ +\ q^{\,2})\ +\ \eta\,v^{\,2}\ +\ b_{\,0}\,\zeta^{\,2}\ +\ 2\,b\,\zeta\,h\ +\ b\,h^{\,2}\Bigr)_{\,t}\ + \\
  \Bigl[\,\zeta\,w^{\,3}\ +\ h\,u\,(u^{\,2}\ +\ q^{\,2})\ +\ \eta\,v^{\,3}\ +\ 2\,p\,\Q\ +\ 2\,b_{\,0}\,\zeta^{\,2}\,w\ +\ 2\,b\,h\,\zeta\,w\ +\ 2\,b\,(\zeta + h)\,h\,u\,\Bigr]_{\,x} \\
  =\ -2\,\bigl(b_{\,0}\,\zeta\,w\ +\ b\,h\,u\bigr)\,d_{\,x}\ -\ \E\ -\ 2\,w\,u_{\,\ast}^{\,2}\,.
\end{multline}
The unknown variables in the combined System \eqref{eq:sys1} -- \eqref{eq:sys7} are $(\zeta,\,h,\,\eta,\,w,\,u,\,v,\,p,\,q,\,b)\,$. The quantity $q^{\,2}$ characterizes the specific energy of small scale motions and it determines the entrainment velocity into the mixing layer. The quantity $\E$ is the turbulent energy dissipation rate. The closure relations adopted in our study are the following:
\begin{align*}
  \chi^{\,\pm}\ &=\ \sigma^{\,\pm}\,q\,, \qquad \sigma^{\,\pm}\ =\ \const\ \geq\ 0\,, \\
  \E\ &=\ \kappa\,\abs{q}^{\,3}\,, \qquad \kappa\ =\ \const\ \geq\ 0\,.
\end{align*}
One can derive two differential consequences of System \eqref{eq:sys1} -- \eqref{eq:sys7}:
\begin{equation}\label{eq:6a}
  b_{\,t}\ +\ u\,b_{\,x}\ =\ \frac{\chi^{\,-}\,(b_{\,0}\ -\ b)\ -\ \chi^{\,+}\,b}{h}\,,
\end{equation}
\begin{equation}\label{eq:6b}
  q_{\,t}\ +\ u\,q_{\,x}\ =\ \frac{\chi^{\,-}\,\bigl((u - w)^{\,2}\ -\ q^{\,2}\ -\ (b_{\,0} - b)\,h\bigr)\ +\ \chi^{\,+}\,\bigl((u - v)^{\,2}\ -\ q^{\,2}\ -\ b\,h\bigr)\ -\ \E}{2\,h\,q}\,.
\end{equation}
From these two equations it follows that System \eqref{eq:sys1} -- \eqref{eq:sys7} possesses a double contact characteristics $\od{x}{t}\ =\ u\,$. The other characteristics coincide with those of the classical three layer shallow water (or \textsc{Saint}-\textsc{Venant}) equations \cite{Liapidevskii2000}.

\begin{remark}
If the mixing processes are not included (\ie~$\sigma^{\,\pm}\ \equiv\ 0$), then Equations \eqref{eq:6a}, \eqref{eq:6b} become simple transport equations for constant values of $b$ and $q$ along the characteristics.
\end{remark}

\begin{remark}
If we take $h\ \equiv\ 0\,$, Equations \eqref{eq:sys1} -- \eqref{eq:sys7} are not contradictory. In this case, the subset of Equations \eqref{eq:sys1} -- \eqref{eq:sys4} form a closed system of two layer shallow water (or \textsc{Saint}-\textsc{Venant}) equations.
\end{remark}

%%% ----------------------------------------------------------------------- %%%

\section{A lyrical digression}

In \cite{Liapidevskii2004} it was shown that the complete System \eqref{eq:sys1} -- \eqref{eq:sys7} of equations (with $\sigma^{\,+}\ \equiv\ \sigma^{\,-}\ =\ \const$) describe correctly the mixing layer formation in super-critical flows over a slope. Here, by super-criticality we understand that on a considered solution all characteristics are real and positive. Physically, it means that no information can propagate in the upstream direction. In our previous paper \cite{Liapidevskii2018} we investigated the influence of the initial flow stages on the front velocity of the density current under the prescribed heavy fluid flux at the channel entrance (left boundary in that study). When the mixing layer approaches the bottom, the entrainment of sediments stops and this layer becomes an underwater turbulent jet. In the same time, the buoyancy in the boundary layer remains constant, \ie~$b_{\,0}\ =\ \const$. System \eqref{eq:sys1} -- \eqref{eq:sys7} is suitable for the simulation of such turbulent jets if we disable the exchanges with the bottom layer:
\begin{equation*}
  \sigma^{\,+}\ =\ \const\ >\ 0\,, \qquad \sigma^{\,-}\ =\ 0\,.
\end{equation*}
In \cite{Liapidevskii2018} it was shown that the density current front velocity in channels with moderate slopes is essentially determined by the mass flux in the bottom layer. The obtained dependence was found to be in good agreement with available experimental data \cite{Britter1980}.

In the present study we consider a special class of flows, where the mass in-flux at the left boundary is absent. In other words, the channel is closed (at least at the left extremity). The motion is initiated by the fluid of density $\rho_{\,0}\,$, initially at rest, which occupies the left part of the channel $0\ \leq\ x\ \leq\ \ell_{\,0}\,$. This description corresponds to classical lock-exchange flows in laboratory conditions. For $x\ >\ \ell_{\,0}$ sediments of the density $\rho_{\,s}$ may be distributed along the slope. They are initially motionless as well. If sediments are absent, we will have a heavy fluid flow along the slope, which propagates over an incline in the form of a thermal. In this case, the density current front experiences first an acceleration phase\footnote{We note that in the case of non-canonical releases on laterally unbounded domains, gravity currents propagating on inclined surfaces may undergo two acceleration phases \cite{Zgheib2016, Zhu2017}.}, and then, a deceleration \cite{Fukushima2000, Maxworthy2007, Maxworthy2010, Dai2013}. In the case where the sediments with buoyancy $m_{\,s}\ =\ \dfrac{(\rho_{\,s}\ -\ \rho_{\,a})}{\rho_{\,a}}\;g\,h_{\,s}$ are distributed all along the slope, then the deceleration phase might be absent. It is in this case that we deal with a self-sustained density gradient flow or an underwater avalanche. The main source of this process lies in the potential energy of sediment deposits.

In every situation mentioned hereinabove, the initial presence of the sediment layer of density $\rho_{\,0}$ at the left compartment ($0\ \leq\ x\ \leq\ \ell_{\,0}$) influences only \emph{initial stages} of the flow formation. In fact, the bottom part of this layer contains an amount heavy fluid, which is quickly entrained into the flow head. Then, this mass goes entirely into the turbulent mixing layer, especially if we neglect the sedimentation velocity. Thus, on relatively large times, in all the cases we obtain a turbulent jet propagating along the bottom further downslope. Finally, the governing equations of motion can be readily deduced from System \eqref{eq:sys1} -- \eqref{eq:sys7}:
\begin{align*}
  h_{\,t}\ +\ \bigl[\,h\,u\,\bigr]_{\,x}\ &=\ \sigma\,q\,, \\
  (b\,h)_{\,t}\ +\ \bigl[\,b\,h\,u\,\bigr]_{\,x}\ &=\ 0\,, \\
  v_{\,t}\ +\ \bigl[\,\half\,v^{\,2}\ +\ p\,\bigr]_{\,x}\ &=\ 0\,,
\end{align*}
\begin{equation*}
  (h\,u\ +\ \eta\,v)_{\,t}\ +\ \Bigl[\,h\,u^{\,2}\ +\ \half\,b\,h^{\,2}\ +\ \H\,p\,\Bigr]_{\,x}\ =\ -\,(p\ +\ b\,h)\,d_{\,x}\ -\ u_{\,\ast}^{\,2}\,,
\end{equation*}
\begin{multline*}
  \bigl(h\,(u^{\,2} + q^{\,2}) + \eta\,v^{\,2} + b\,h^{\,2}\bigr)_{\,t}\ +\ \bigl[\,h\,u\,(u^{\,2} + q^{\,2}) + \eta\,v^{\,3} + 2\,p\,\Q + 2\,b\,h^{\,2}\,u\,\bigr]_{\,x}\\
   =\ -\,2\,b\,h\,u\,d_{\,x}\ -\ \kappa\,\abs{q}^{\,3}\ -\ 2\,u\,u_{\,\ast}^{\,2}\,,
\end{multline*}
where $\H\ \eqdef\ h\ +\ \eta\ \equiv\ \H_{\,0}\ -\ d\,(x)\,$, $\Q\ =\ h\,u\ +\ \eta\,v$ and $\sigma\ \equiv\ \sigma^{\,+}\,$. Note that the friction terms in two last equations appear due to the absence of the bottom \emph{slipping} layer in the density flows under consideration.

%%% ----------------------------------------------------------------------- %%%

\section{Nomenclature}
\label{app:nom}

In the main text above we used the following notations (this list is not exhaustive):
\begin{description}
  \item[$\equiv\ $] equal identically
  \item[$\gtrsim\ $] approximatively greater
  \item[$\zeta\ $] sediments layer thickness
  \item[$b\ $] buoyancy
  \item[$d\,(x)\ $] bathymetry data (constant inclined bottom in our study)
  \item[$\H\,(x)\ $] total channel depth
  \item[$p\ $] fluid pressure
  \item[$m\ $] scaled mass, \ie~$m\ =\ b\cdot\zeta$
  \item[$m_{\,s}\ $] sediments scaled mass
  \item[$m_{\,l}\ $] heavy fluid mass
  \item[$h_{\,j}\ $] total depth of the layer $j$
  \item[$q\ $] depth-averaged turbulent fluctuating velocity component in the mixing layer
  \item[$\rho_{\,j}\ $] fluid density in the layer $j$
  \item[$\rho_{\,s}\ $] density of sediments
  \item[$\rho_{\,0}\ $] heavy fluid density (constant)
  \item[$\rho_{\,a}\ $] ambient fluid density (constant)
  \item[$\phi\ $] bottom slope angle
  \item[$\alpha\ $] bottom slope, \ie~$\alpha\ =\ \tan\phi$
  \item[$\D\ $] dimensional front velocity
  \item[$\D_{\,\mathrm{f}}\ $] dimensionless front velocity
  \item[$t\ $] time variable
  \item[$x\ $] ``horizontal'' coordinate along the bottom slope
  \item[$y\ $] ``vertical'' coordinate, normal to the bottom
  \item[$\tilde{g}\ $] reduced gravity acceleration
  \item[$\chi\ $] entrainment rate
  \item[$\nu\ $] wave characteristics speed in hyperbolic equations
  \item[$u_{\,\ast}\ $] friction velocity at the solid bottom
  \item[$\Fr\ $] the dimensionless \textsc{Froude} number
  \item[$\xi\ $] self-similar variable, \eg~$\xi\ =\ \frac{x}{t}$
  \item[$\vtheta\ $] similarity exponent, \eg~$\xi\ =\ \frac{x}{t^{\,\vtheta\,+\,1}}$
  \item[$\lambda\ $] travelling wave slope
  \item[$\eta\ $] ambient fluid layer thickness
  \item[$v\ $] depth-averaged velocity of the ambient fluid
  \item[$\ell_{\,0}\ $] lock height in lock-exchange experiments
  \item[$L\ $] computational domain length
  \item[$\kappa\ $] energy dissipation coefficient
\end{description}

For conservation laws we employ the following notation:
\begin{equation*}
  (V)_{\,t}\ +\ \bigl[\,F\,\bigr]_{\,x}\ =\ S\,,
\end{equation*}
where $V$ is the conserved quantity, $F$ is the corresponding flux and $S$ is the source term.

%%% ----------------------------------------------------------------------- %%%

%%% Bibliography
\addcontentsline{toc}{section}{References}
\bibliographystyle{abbrv}
\bibliography{biblio}
\bigskip\bigskip

\end{document}